\documentclass[useAMS,usenatbib]{mn2e}
\usepackage{epsfig}
\usepackage{amssymb}
\usepackage{amsmath}
\usepackage{lscape}
\usepackage{longtable}
\usepackage[normal]{caption}
\usepackage{appendix}

\def\arcsec{$^{\prime\prime}$}

\title[The \textit{Herschel} census of infrared SEDs through cosmic time]{The \textit{Herschel} census of infrared SEDs
through cosmic time\thanks{{\it Herschel} is an ESA space observatory with science instruments provided by European-led Principal Investigator consortia and with important participation from NASA.}}

\author[M.~Symeonidis et al.]
{\parbox{\textwidth}{\raggedright M.~Symeonidis,$^{1}$\thanks{E-mail: \texttt{msy@mssl.ucl.ac.uk}}
M.~Vaccari,$^{2}$
S.~Berta,$^{3}$
M.J.~Page,$^{1}$
D.~Lutz,$^{3}$
V. ~Arumugam,$^{17}$
H.~Aussel,$^{4}$
J.~Bock,$^{5,6}$
A. ~Boselli,$^{4}$
V.~Buat,$^{7}$
P. L. Capak,$^{5}$
D.L.~Clements,$^{8}$
A. ~Conley,$^{26}$
L. ~Conversi,$^{27}$
A.~Cooray,$^{9,5}$
C.D. ~Dowell,$^{5,6}$
D.~Farrah,$^{36}$
A.~Franceschini,$^{2}$
E.~Giovannoli,$^{39}$
J. ~Glenn,$^{26,28}$
M. ~Griffin,$^{29}$
E.~Hatziminaoglou,$^{11}$
\mbox{H.-S.}~Hwang,$^{40}$
E. ~Ibar,$^{30}$
O. ~Ilbert,$^{23}$
R.J. ~Ivison,$^{17,30}$
E.~Le Floc'h,$^{4}$
S.~Lilly,$^{21}$
J. S. ~Kartaltepe,$^{19}$
B.~Magnelli,$^{3}$
G.~Magdis,$^{13}$
L. ~Marchetti,$^{2}$
H.T. ~Nguyen,$^{5,6}$
R.~Nordon,$^{37}$
B.~O'Halloran,$^{8}$
S.J.~Oliver,$^{10}$
A.~ Omont,$^{31}$
A.~ Papageorgiou,$^{29}$
H.~Patel,$^{8}$
C.P.~Pearson,$^{14,15}$
I. ~P\'{e}rez-Fournon,$^{32,33}$
M. ~Pohlen,$^{29}$
P.~Popesso,$^{3}$
F.~Pozzi,$^{16}$
D. ~Rigopoulou,$^{13,14}$
L.~Riguccini,$^{38}$
D.~Rosario,$^{3}$
I.G.~Roseboom,$^{17}$ 
M. Rowan-Robinson,$^{8}$ 
M. ~Salvato,$^{22}$
B. ~Schulz,$^{5,34}$
Douglas Scott,$^{35}$
N. ~Seymour,$^{18}$ 
D.L. ~Shupe,$^{5,34}$ 
A.J. Smith,$^{10}$
I. ~Valtchanov,$^{27}$ 
L. ~Wang,$^{10}$
C.K. ~Xu,$^{5,34}$ 
M. ~Zemcov,$^{5,6}$  and
S.~Wuyts$^{3}$}\vspace{0.4cm}\\
\parbox{\textwidth}{\raggedright $^{1}$Mullard Space Science Laboratory, University College London, Holmbury St. Mary, Dorking, Surrey RH5 6NT, UK\\
$^{2}$Dipartimento di Astronomia, Universit\`{a} di Padova, vicolo Osservatorio, 3, 35122 Padova, Italy\\
$^{3}$Max-Planck-Institut f\"ur Extraterrestrische Physik (MPE), Postfach 1312, 85741, Garching, Germany\\
$^{4}$Laboratoire AIM-Paris-Saclay, CEA/DSM/Irfu - CNRS - Universit\'e
Paris Diderot, CE-Saclay, pt courrier 131, F-91191 Gif-sur-Yvette, France\\
$^{5}$California Institute of Technology, 1200 E. California Blvd., Pasadena, CA 91125, USA\\
$^{6}$Jet Propulsion Laboratory, 4800 Oak Grove Drive, Pasadena, CA
91109, USA\\
$^{7}$Laboratoire d'Astrophysique de Marseille, OAMP, Universit\'e
Aix-marseille, CNRS, 38 rue Fr\'ed\'eric Joliot-Curie, 13388 Marseille
cedex 13, France\\
$^{8}$Astrophysics Group, Imperial College London, Blackett Laboratory, Prince Consort Road, London SW7 2AZ, UK\\
$^{9}$Dept. of Physics \& Astronomy, University of California, Irvine, CA 92697, USA\\
$^{10}$Astronomy Centre, Dept. of Physics \& Astronomy, University of Sussex, Brighton BN1 9QH, UK\\
$^{11}$ESO, Karl-Schwarzschild-Str. 2, 85748 Garching bei M\"unchen, Germany\\
$^{13}$Department of Astrophysics, Denys Wilkinson Building, University of Oxford, Keble Road, Oxford OX1 3RH, UK\\
$^{14}$RAL Space, Rutherford Appleton Laboratory, Chilton, Didcot, Oxfordshire OX11 0QX, UK\\
$^{15}$Institute for Space Imaging Science, University of Lethbridge, Lethbridge, Alberta, T1K 3M4, Canada\\
$^{16}$INAF-Osservatorio Astronomico di Roma, via di Franscati 33, 00040 Monte Porzio Catone, Italy\\
$^{17}$Institute for Astronomy, Blackford Hill, Edinburgh EH9 3HJ, UK\\
$^{18}$CSIRO Astronomy \& Space Science, PO Box 76, Epping, NSW 1710, Australia\\
$^{19}$ National Optical Astronomy Observatory, 950 N. Cherry Ave, Tucson, AZ 85719, US\\
$^{21}$ Institute for Astronomy, Wolfgang-Pauli-Strasse 27, 8093 Zurich, Switzerland\\
$^{22}$ Max-Planck-Institut fur extraterrestrische Physik, Giessenbachstrasse 1, 85748 Garching, Germany\\
$^{23}$ Laboratoire d’Astrophysique de Marseille, Universite de Provence, CNRS, BP 8, Traverse du Siphon, 13376 Marseille Cedex 12, France\\ 
$^{26}$ Center for Astrophysics and Space Astronomy 389-UCB, University of Colorado, Boulder, CO 80309, USA\\
$^{27}$ Herschel Science Centre, European Space Astronomy Centre, Villanueva de la Ca\~nada, 28691 Madrid, Spain\\
$^{28}$ Dept. of Astrophysical and Planetary Sciences, CASA 389-UCB, University of Colorado, Boulder, CO 80309, USA\\
$^{29}$ School of Physics and Astronomy, Cardiff University, Queens Buildings, The Parade, Cardiff CF24 3AA, UK\\
$^{30}$ UK Astronomy Technology Centre, Royal Observatory, Blackford Hill, Edinburgh EH9 3HJ, UK \\
$^{31}$ Institut d'Astrophysique de Paris, UMR 7095, CNRS, UPMC Univ. Paris 06, 98bis boulevard Arago, F-75014 Paris, France \\
$^{32}$ Instituto de Astrof\'{i}sica de Canarias (IAC), E-38200 La Laguna, Tenerife, Spain \\
$^{33}$ Departamento de Astrof\'{i}sica, Universidad de La Laguna (ULL), E-38205 La Laguna, Tenerife, Spain \\
$^{34}$ Infrared Processing and Analysis Center, MS 100-22, California Institute of Technology, JPL, Pasadena, CA 91125, USA\\
$^{35}$ Department of Physics $\&$ Astronomy, University of British
Columbia, 6224 Agricultural Road, Vancouver, BC V6T 1Z1, Canada\\
$^{36}$ Department of Physics, Virginia Tech, Blacksburg, VA 24061,
USA\\
$^{37}$ School of Physics and Astronomy, The Raymond and Beverly
Sackler Faculty of Exact Sciences, Tel-Aviv University, Tel-Aviv
69978, Israel\\
$^{38}$ Astrophysics Branch, NASA/Ames Research Center, MS 245-6,
Moffett Field, CA 94035\\
$^{39}$ Physics Department, University of the Western Cape, Private
Bag X17, 7535, Bellville, Cape Town, South Africa\\
$^{40}$ Smithsonian Astrophysical Observatory, 60 Garden Street, Cambridge, MA 02138, USA}}

\begin{document}

\date{Accepted  Received; in original form}

\pagerange{\pageref{firstpage}--\pageref{lastpage}} \pubyear{2010}

\maketitle

\label{firstpage}

\begin{abstract}

Using \textit{Herschel} data from the deepest SPIRE and PACS surveys
(HerMES and PEP)  in COSMOS, GOODS-S
and GOODS-N, we examine the dust properties of infrared (IR)-luminous
($L_{\rm IR}$$>$10$^{10}$\,L$_{\odot}$) galaxies at $0.1<z<2$ and determine how these evolve with cosmic time. The unique angle of this work is the rigorous
analysis of survey selection effects, making this the first
study of the star-formation-dominated, IR-luminous
population within a framework almost entirely free of selection
biases. We find that IR-luminous galaxies have spectral energy
distributions (SEDs) with broad far-IR peaks
characterised by cool/extended dust emission and average dust temperatures in the
25--45\,K range. Hot ($T>45$) SEDs and cold ($T<25$), cirrus-dominated
SEDs are rare, with most sources being within the range occupied by
warm starbursts such as M82 and cool spirals such as M51. We observe a luminosity-temperature ($L-T$)
relation, where the average dust temperature of log\,[$L_{\rm
  IR}$/L$_{\odot}$]=12.5 galaxies is about 10\,K higher than that of their
log\,[$L_{\rm IR}$/L$_{\odot}$]=10.5 counterparts. However, although the increased
dust heating in more luminous systems is the driving factor behind the
$L-T$ relation, the increase in dust mass and/or starburst size with luminosity plays a
dominant role in shaping it. Our results show that the dust conditions
in IR-luminous sources evolve with cosmic time: at high redshift,
dust temperatures are on average up to 10\,K lower than what
is measured locally ($z\lesssim0.1$). This is manifested as a flattening of the
$L-T$ relation, suggesting that (U)LIRGs in the early
Universe are typically characterised by a more extended dust
distribution and/or higher dust masses than local equivalent sources. Interestingly, the evolution in dust temperature
is luminosity dependent, with the fraction of LIRGs with $T<$35\,K 
showing a 2-fold increase from $z\sim$0 to $z\sim$2, whereas that of
ULIRGs with $T<$35\,K shows a
6-fold increase. Our results suggest a greater diversity in the
IR-luminous population at high redshift, particularly for ULIRGs.
\end{abstract}


\section{Introduction}
\label{sec:introduction}

The discovery of a class of infrared (IR)-luminous ($L_{\rm IR}$\,$>$\,10$^{10}$\,L$_{\odot}$) galaxies in the
60s (e.g. Johnson 1966\nocite{Johnson66}; Low $\&$ Tucker
1968\nocite{LT68}; Kleinmann $\&$ Low 1970\nocite{KL70}), followed by the detection of the cosmic infrared background
(Puget et al. 1998\nocite{Puget96}; Hauser et al. 1998\nocite{Hauser98}) unfolded a new era in extragalactic
astronomy. Infrared/submm surveys with \textit{IRAS}
(Neugebauer et al. 1984\nocite{Neugebauer84}), \textit{ISO} (Kessler
et al. 1996\nocite{Kessler96}), JCMT/SCUBA (Holland et al. 1999\nocite{Holland99}),
\textit{Spitzer} (Werner et al. 2004\nocite{Werner04}) and \textit{AKARI} (Murakami et al. 2007\nocite{Murakami07}), revealed that the early Universe was more active than
previously thought, uncovering a large number of dust-enshrouded galaxies whose
bolometric luminosity emerges almost entirely in the infrared (e.g. Soifer
et al. 1984a\nocite{Soifer84a}; Sanders $\&$ Mirabel
1996\nocite{SM96}; Lutz et al. 1996\nocite{Lutz96};
2005\nocite{Lutz05}; Rowan-Robinson et
al. 1997\nocite{RR97}; 2005\nocite{RR05}; Lisenfeld, Isaak $\&$ Hills
2000\nocite{LIH00}; Goto et al. 2010\nocite{Goto10} and many
more). These IR-luminous galaxies are rare in the local Universe
(e.g. Kim $\&$ Sanders 1998\nocite{KS98}) but
exhibit a strong increase in number density at earlier
epochs (e.g. Takeuchi, Buat $\&$ Burgarella 2005\nocite{TBB05}), being responsible for about half the total light emitted from
all galaxies integrated through cosmic time (e.g. Gispert et al. 2000\nocite{GLP00};
Lagache et al. 2005\nocite{LPD05}; Dole et
al. 2006\nocite{Dole06}). The abundance of these sources at high redshifts ($z\sim$1--4; e.g. Hughes et
al. 1998\nocite{Hughes98}; Eales et al. 1999\nocite{Eales99},
2000\nocite{Eales00}; Blain et al. 2004\nocite{Blain04a}; Le Floc'h et
al. 2004\nocite{LeFloch04}; Schinnerer et al. 2008\nocite{Schinnerer08}; Pannella et al. 2009\nocite{Pannella09}) indicates that they started
forming very early in cosmic history, potentially challenging the
hierarchical paradigm of $\Lambda$CDM (e.g. Granato et
al. 2004\nocite{Granato04}; Baugh et al. 2005\nocite{Baugh05}; Bower et
al. 2006\nocite{Bower06}; Somerville et al. 2008\nocite{Somerville08}).

IR-luminous galaxies are the ideal laboratories for studies of
galaxy formation and evolution through chemical enrichment,
star-formation, black hole accretion and stellar mass
build-up. They hide an immensely active
interstellar medium (ISM; e.g. Lutz et al. 1998\nocite{Lutz98}; Farrah
et al. 2003\nocite{Farrah03}; Narayanan et al. 2005\nocite{Narayanan05}; Sturm et al. 2010\nocite{Sturm10}) and are the ultimate stellar nurseries, with star-formation rates (SFRs) up to a few
thousand times higher than Milky Way (MW)-type galaxies (e.g. Kennicutt
1998\nocite{Kennicutt98}; Egami et al. 2004\nocite{Egami04}; Choi et al. 2006\nocite{Choi06}; Rieke et al. 2009\nocite{Rieke09}). In addition, they are amongst
the most massive galaxies in the Universe (e.g. Dye et
al. 2008\nocite{Dye08}; Micha{\l}owski et al. 2010\nocite{MHW10}) and often their morphologies
show signs of interactions and mergers (e.g. Sanders $\&$ Mirabel
1996\nocite{SM96}; Farrah et al. 2001\nocite{Farrah01};
2002\nocite{Farrah02}; Moustakas et al. 2004\nocite{Moustakas04};
Kartaltepe et al. 2010b\nocite{Kartaltepe10b}). Finally, they
frequently harbour an active galactic nucleus (AGN), which is commonly considered a key player in the
evolution of the system (e.g. Genzel et al. 1998\nocite{Genzel98}; Ptak et al. 2003\nocite{Ptak03}; Alexander et al. 2005\nocite{Alexander05a};
Page et al. 2012\nocite{Page12}). 

Until recently our view of the IR-luminous galaxy population
at high redshift has been based on data from the space observatories
\textit{ISO} and \textit{Spitzer}, as well as ground-based submm/mm
facilities such as JCMT/SCUBA, APEX/LABOCA, IRAM/MAMBO and SMA/AzTEC. 
Although huge advances have been made in our understanding of the
nature and evolution of these sources, it has been challenging to reconcile the
data from space observatories with comparable ground-based
IR/mm datasets. In recent years it has
become increasingly apparent that, besides strong evolution in IR galaxy
number density (e.g. Le Floc'h et al. 2005\nocite{LeFloch05}; Huynh et
al. 2007\nocite{Huynh07b}; Magnelli et al. 2009, Berta et al. 2010;
2011), the physical properties of IR galaxies might also
evolve with time, with the rate of evolution potentially changing as a function of luminosity (Seymour et al. 2010\nocite{Seymour10}). 
Studies of the local Universe showed
that ultraluminous infrared galaxies (ULIRGs) are characterised by warm average dust temperatures
(e.g. Soifer et al. 1984\nocite{Soifer84b}; Klaas et
al. 1997\nocite{Klaas97}; Clements, Dunne $\&$ Eales 2010\nocite{CDE10}), strong
silicate absorption and PAH emission features in their mid-IR continuum (e.g Brandl et al. 2006; Armus et al.
2007), as well as compact starburst sizes (e.g. Condon et al. 1991\nocite{Condon91};
Soifer et al. 2001\nocite{Soifer01}). However, with the onset of submm/mm
facilities which probed the early Universe ($z>$2), such as SCUBA in the late
1990s, a different picture emerged. Many IR-luminous
galaxies at high redshift were found to be less compact than their local counterparts
(e.g. Tacconi et al. 2006\nocite{Tacconi06}; Iono et
al. 2009\nocite{Iono09}; Rujopakarn et al. 2011), exhibiting stronger PAH emission (Farrah et
al. 2007\nocite{Farrah07}; 2008\nocite{Farrah08}; Valiante et al. 2007\nocite{Valiante07}) and a greater abundance of cold dust (Kov\'acs et al. 2006\nocite{Kovacs06}; Pope et
al. 2006\nocite{Pope06}; Coppin
et al. 2008\nocite{Coppin08}). It was later shown that these differences in the
measured dust properties were partly
due to selection effects and partly due to evolution (Symeonidis et
al. 2009; 2011a\nocite{Symeonidis11a}). Moreover, the exploitation of long wavelength data from \textit{ISO}
(Rowan-Robinson et al. 2005\nocite{RR05}) and
\textit{Spitzer} (Symeonidis et al. 2007\nocite{Symeonidis07}; 2008\nocite{Symeonidis08}), enabled the discovery of
IR-luminous galaxies at $<$1, with a spectrum of properties which
overlapped with both the local population and the SCUBA-detected,
$z>$2, sources, providing the missing link between the two (Symeonidis et
al. 2009). 

The launch of the \textit{Herschel Space Observatory}\footnote{\textit{Herschel} is an ESA space observatory with science instruments provided by European-led
  Principal Investigator consortia and with important participation
  from NASA.} (Pilbratt et al. 2010\nocite{Pilbratt10}) has opened a
new window in infrared astronomy, as it is the only facility to date and for the
foreseeable future to perfectly span the wavelength range in which
most of the Universe's obscured radiation emerges (70-500\,$\mu$m), uncovering
unprecedented numbers of dust-enshrouded galaxies over a sizeable
fraction of cosmic time. The large dynamical range of the PACS
(Poglitsch et al. 2010\nocite{Poglitsch10}) and SPIRE (Griffin et al. 2010\nocite{Griffin10})
instruments, have enabled spectral energy distributions (SEDs) to be compiled
for a large range of objects, both for AGN (e.g. Hatziminaoglou et al. 2010\nocite{Hatziminaoglou10}; Seymour et al. 2011\nocite{Seymour11}) and
star-forming galaxies (e.g. Rowan-Robinson et al. 2010\nocite{RR10}).  Recent studies of the properties
of the IR-luminous galaxy population using \textit{Herschel} data
provide an excellent showcase of the
capabilities of this observatory (some examples from the multitude of
\textit{Herschel} papers on this topic: Amblard et al. 2010\nocite{Amblard10}; Rowan-Robinson et
al. 2010\nocite{RR10}; Magnelli et al. 2010\nocite{Magnelli10}; 2012\nocite{Magnelli12}, Magdis
et al. 2010\nocite{Magdis10}; 2011\nocite{Magdis10}; Gruppioni et
al. 2010\nocite{Gruppioni10}; Eales et al. 2010b\nocite{Eales10b}; Hwang et
al. 2010\nocite{Hwang10}; Elbaz et al. 2010\nocite{Elbaz10};
2011\nocite{Elbaz11}; Rodighiero et al. 2010\nocite{Rodighiero10}; Ivison et
al. 2010\nocite{Ivison10}; Buat et al. 2010\nocite{Buat10}; Dye et
al. 2010\nocite{Dye10}; Berta et
al. 2010\nocite{Berta10}; 2011\nocite{Berta11}; Dunne et
al. 2011\nocite{Dunne11}; Symeonidis et
al. 2011b\nocite{Symeonidis11b}; Kartaltepe et al. 2012\nocite{Kartaltepe12}). Results from these studies carried
out during the Science Demonstration Phase (SDP) of the largest
extragalactic surveys, HerMES (\textit{Herschel} multi-tiered
extragalactic survey; Oliver et al. 2012\nocite{Oliver12}), PEP (PACS
Evolutionary Probe; Lutz et al. 2011\nocite{Lutz11}) and H-ATLAS (\textit{Herschel}
Astrophysical Terahertz Large Area Survey; Eales et
al. 2010a\nocite{Eales10a}), confirmed previous findings on the
diversity of IR-luminous galaxy properties at high redshift, as well
as the existence of high-redshift sources with no local equivalents, moving us closer in understanding the complex nature
of the infrared galaxy population up to $z\sim$3. 

In this paper we report a comprehensive study of the SEDs and dust
temperatures of IR-luminous galaxies up to $z \sim$2, using the deepest available \textit{Herschel}/PACS and SPIRE data acquired as part
of the PEP and HerMES consortia in the COSMOS and GOODS (N $\&$ S) fields. A key aspect of our work is the attempt to eradicate
survey selection effects, an issue which has plagued previous attempts to
canvas the range of infrared SEDs (see Symeonidis et al. 2011a). Thus for
the first time we are able to examine the aggregate properties (infrared luminosity, dust
temperature, SED shape) of IR galaxies within an almost entirely bias-free framework. The paper is laid out as follows: the introduction is
followed by a section on the sample selection (section
\ref{sec:sample}) and SED measurements (section
\ref{sec:measurements}). In section \ref{sec:AGN} we discuss how we
deal with AGN, in order to obtain a sample which is star-formation
dominated in the infrared. Section \ref{sec:selection} is 
devoted to treatment of selection effects, enabling us to assemble a complete sample of IR
galaxies. In section \ref{sec:results} we present our results and
discuss them in section \ref{sec:discussion}. Finally our summary and conclusions
are presented in section \ref{sec:conclusions}. Throughout we employ a
concordance $\Lambda$CDM cosmology of H$_0$=70\,km\,s$^{-1}$Mpc$^{-1}$,
$\Omega_{\rm M}$=1-$\Omega_{\rm \Lambda}$=0.3.

\section{The \textit{Herschel} sample}
\label{sec:sample}

\subsection{Initial selection}
\label{sec:initial_selection}
The starting point for this work are data from \textit{Herschel}, covering three
extragalactic fields: the Great
Observatories Origins Deep Survey (GOODS)-North and South (Giavalisco et al. 2002\nocite{Giavalisco04}) and
the Cosmic Evolution Survey (COSMOS) field (Scoville et
al. 2007\nocite{Scoville07}). We use PACS 100 and 160\,$\mu$m
and SPIRE 250, 350 and
500\,$\mu$m images, acquired as part of PEP and HerMES respectively. 
Source extraction from the PACS and
SPIRE images\footnote{The data presented in
  this paper will be released through the {\em Herschel} Database in
  Marseille HeDaM ({hedam.oamp.fr/HerMES})} is performed on the IRAC-3.6\,$\mu$m
positions of the f$_{24}$$\ge$30\,$\mu$Jy GOODS (N and S) sources and
f$_{24}$$\ge$60\,$\mu$Jy COSMOS sources, as described in Magnelli et
al. (2009\nocite{Magnelli09}) and Roseboom et
al. (2010\nocite{Roseboom10}; 2012\nocite{Roseboom12}). This method of
source extraction on prior positions is widely used and
enables identifications of secure counterparts over the whole SED. In this case however, its significant
advantage, lies in its ability to effectively deal with source
blending in the \textit{Herschel} bands, particularly for
SPIRE where the beam is large (18.1, 24.9 and 36.6 arcsec FWHM at 250, 350 and
500\,$\mu$m respectively; Nguyen et al. 2010). By using prior
information to identify galaxies in the \textit{Herschel} images, we are able to
extract `clean' photometry for each galaxy, even for those which appear blended in the PACS and SPIRE bands.
For information on the
GOODS \textit{Spitzer}/MIPS 24\,$\mu$m data see Magnelli et al. (2009\nocite{Magnelli09})
and for information on the  COSMOS \textit{Spitzer}/MIPS 24\,$\mu$m data see Sanders et al. (2007\nocite{Sanders07}); Le Floc'h et
al. (2009\nocite{LeFloch09}). 
The 3\,$\sigma$ sensitivity limits of the PACS 100 and 160\,$\mu$m
catalogues respectively are 5 and
10\,mJy for COSMOS, 3 and 6\,mJy for GOODS-N and 1 and 2\,mJy for
GOODS-S. A 3\,$\sigma$ detection in
SPIRE using prior positions and the cross-identification method of
Roseboom et al. (2010) is approximately 8, 11 and 13\,mJy at 250, 350
and 500\,$\mu$m in all fields. In the case of the PACS bands $\sigma$ is only the photometric
error, whereas for the SPIRE bands, $\sigma$ includes confusion
error  (see Nguyen et al. 2010\nocite{Nguyen10} for the SPIRE
confusion limits).

\begin{table*}
\centering
\caption{Table showing the detection statistics of the \textit{Herschel}
  sample used in this work. The first
  and second columns show the field and number of 24\,$\mu$m sources
  with $f_{24}$$>$30\,$\mu$Jy and $f_{24}$$>$60\,$\mu$Jy for GOODS (N
  $\&$ S) and COSMOS respectively,
  whose positions are used as priors for source extraction in the
  \textit{Herschel} bands. Column 3 shows the total number of
  `isolated' 24\,$\mu$m sources defined as having no companion within
  8\arcsec; the percentage in brackets is calculated out of the number in column
  2. Column 4 shows the fraction detected at 100 and 160\,$\mu$m,
  whereas column 5 shows the fraction detected
  at 160 and 250\,$\mu$m. The final column shows what fraction of the 24\,$\mu$m
  population is detected when the two criteria are used in
  disjunction, i.e. [100 and 160\,$\mu$m] OR [160 and
  250\,$\mu$m]. This is the criterion used to select the initial
  \textit{Herschel} sample (section \ref{sec:initial_selection}). The fractions shown
  in columns 4, 5 and 6 are out of the number of sources in column
  3. As also mentioned in section \ref{sec:initial_selection}, the 3\,$\sigma$ sensitivity limits of the PACS 100 and 160\,$\mu$m
catalogues respectively are 5 and
10\,mJy for COSMOS, 3 and 6\,mJy for GOODS-N and 1 and 2\,mJy for
GOODS-S. A 3\,$\sigma$ detection in
SPIRE using prior positions and the cross-identification method of
Roseboom et al. (2010) is approximately 8, 11 and 13\,mJy at 250, 350
and 500\,$\mu$m in all fields. }
\begin{tabular}{l|l|l|l|l|l|}
\hline 
field&number of      & number of        & fraction detected  at & fraction
detected at  & fraction detected at \\ 
       &24$\mu$m sources & `isolated' 24\,$\mu$m sources &100+160\,$\mu$m ($>$3$\sigma$) &
      160+250\,$\mu$m ($>$3$\sigma$) &[100+160\,$\mu$m] OR [160+250\,$\mu$m]\\
(1) &(2) &(3)&(4)&(5)&(6)\\
\hline
GOODS-N&2149& 1401 (65$\%$)&7$\%$&7$\%$&9$\%$\\
GOODS-S& 2252& 1580 (70$\%$)&21$\%$&9$\%$&22$\%$\\
COSMOS &52092&33407 (64$\%$) &5$\%$&6$\%$&7$\%$\\
\hline
\end{tabular}
\label{table:detection_rates}
\end{table*}

The initial selection for our sample includes all 24\,$\mu$m sources
that have detections (at least 3$\sigma$) at [100 and
160\,$\mu$m] OR [160 and 250\,$\mu$m] (where `OR' is the operator
representing disjunction in Boolean logic). 
This ensures that (i) the sample consists of IR-luminous galaxies ($L_{\rm IR}$\,$>$\,10$^{10}$\,L$_{\odot}$),
(ii) the sample is as complete as possible over a large redshift range with respect
to SED types, given the PACS and
SPIRE selection functions (see section \ref{sec:selection} for more
details) and (iii) there are at least 3 reliable photometric points in the SED (24\,$\mu$m+ 2
\textit{Herschel} bands) for subsequent measurements. 

Our selection is in essence a colour selection rather than a single
band selection, as we require sources to be detected at both 24
and 160\,$\mu$m (the additional \textit{Herschel} photometry at
100 or 250\,$\mu$m has a small effect on the sample selection but enables more
accurate SED measurements; see section \ref{sec:selection}). As a result, the detection rate is more strongly
dependent on the SED shape, in our case the mid-to-far-IR continuum slope, than for
a typical flux limited survey. Given the flux limits reported earlier, the 24\,$\mu$m survey is 66, 166 and 200 times
deeper than the PACS 160\,$\mu$m survey in GOODS-S,
COSMOS and GOODS-N respectively. We find that the different ratios in
flux density limits ($f_{160}^{\rm lim}/f_{24}^{\rm lim}$)
between the 3 fields, introduce a bias with respect to the SED shapes
that are detected in each survey particularly for objects with flux
densities close to the limit. To mitigate this effect, we match the relative PACS-24\,$\mu$m
depths of the GOODS fields to the COSMOS survey, as the latter covers the
largest area and hence dominates the statistics of the final
sample. This is done as follows: the GOODS-S sample is
cut at $f_{160}$=5\,mJy and the GOODS-N sample
is cut at $f_{24}$=36\,$\mu$Jy, so that $f_{160}^{\rm lim}/f_{24}^{\rm
  lim} \sim166$
in all cases. Note that matching the samples in this
way ensures that the \emph{relative} biases between the surveys are minimised, i.e. that all three
surveys probe the same range of SED types, however it does not deal
with \emph{absolute} biases; these are dealt with in section
\ref{sec:selection}. 

Besides the photometric selection criteria,
we also restrict the sample to sources which have no other 24\,$\mu$m companions within 8\arcsec; i.e. `isolated' sources. This allows us to work with more reliable photometry, as at
the longer wavelengths, where the \textit{Herschel} beam is large,
flux extraction in the \textit{Herschel} bands can be problematic when
dealing with blended sources. The choice of an 8\arcsec radius is
larger than the
24\,$\mu$m beam (6\arcsec) and visual inspection shows that it is
sufficient to eliminate problematic blends. In addition, 8\arcsec is the
scale of the first airy ring of bright 24\,$\mu$m sources. In cases
where there is a companion within the first airy ring, the
\textit{Herschel} flux is assigned to the bright source in the centre, if the companion
source is 10 or more times fainter (Roseboom et al. 2010). By
eliminating such cases from our sample, we avoid the occurance of a potential
bias causing the measured \textit{Herschel} flux
density to positively correlate with the 24\,$\mu$m flux density.
As a result, for the 24\,$\mu$m sources we subsequently use, there have been no prior
assumptions when assigning \textit{Herschel} fluxes. This is confirmed
by performing a K-S test on the flux density distribution of the whole
24\,$\mu$m population and that of the `isolated' 24\,$\mu$m sources in
each field. We find the two to be entirely
consistent, suggesting that our approach works well in eliminating
problematic sources without introducing systematic biases. 

At this point, our assembled \textit{Herschel} sample consists of 2500 sources (2206 from COSMOS, 173 from
GOODS-S and 121 from GOODS-N). Some statistics for
the initial sample are shown in table \ref{table:detection_rates}.

\subsection{Redshifts}
\label{sec:redshifts}
The redshifts we use are a combination of spectroscopic and
photometric, assembled from various catalogues: Berta et al. (2011\nocite{Berta11}) for GOODS-N, Cardamone et
al. (2010\nocite{Cardamone10}) and Santini et al. (2009\nocite{Santini09}) for GOODS-S and Ilbert et al. (2009\nocite{Ilbert09}) and Salvato et al. (2009);
(2011)\nocite{Salvato09}\nocite{Salvato11} for COSMOS. The optical positions of sources in these catalogues are cross-matched to
the 24\,$\mu$m positions within 1\arcsec. The excellent photometric coverage of these fields and high quality
photometric redshifts available, result in $>$90 per cent of the
sources in our sample having a usable redshift (25 per cent
spectroscopic; although for the final sample about 1/3 are
spectroscopic), leaving 2313 \textit{Herschel} sources for subsequent analysis.
For more details on the quality of photometric redshifts see Appendix
\ref{appendixA}, where we also show that the use of photometric
redshifts does not bias our results.

\section{Measurements}
\label{sec:measurements}
We fit the photometry of the \textit{Herschel} sample, with the Siebenmorgen $\&$
Kr{\"u}gel (2007, hereafter SK07\nocite{SK07}) library of 7208 models, built on the formulation of Kr\"ugel $\&$
Siebenmorgen (1994). This library of templates ranges in 5 free parameters, in
physically acceptable combinations and the shape of each template is
determined by the combination of parameters which define it:
\begin{itemize}
\item the radius of the IR emitting starburst region ($R$), taking discrete
 values of 0.35, 1, 3, 9, 15 kpc
\item the total luminosity of the system ($L_{\rm tot}$) ranging from 10$^{10}$ to
  10$^{15}$\,L$_{\odot}$
\item the visual extinction from the edge to the centre of the
 starburst, taking discrete values of 2.2, 4.5, 7, 9, 18, 35, 70 and 120 mag
\item the ratio of the luminosity of OB stars with hot spots to the total
luminosity ($L_{\rm OB}$/$L_{\rm tot}$) taking discrete values of 40,
60 and 90 per cent for the $\le$3 kpc models and 100 per cent for the 9 and 15kpc models
\item the dust density in the hot spots in units of hydrogen number
 density (cm$^{-3}$) ranging from 100 to 10000, in discrete steps
\end{itemize}

As mentioned earlier, 6-band photometry is available --- 24\,$\mu$m from \textit{Spitzer}/MIPS, 100, 160 from
\textit{Herschel}/ PACS and 250, 350, 500$\mu$m from
\textit{Herschel}/SPIRE --- and at least 3 bands, always
including the 24 and 160\,$\mu$m data, are used in the fitting. 
The normalisation of the templates is varied in order to minimise
$\chi ^2$. The 0.68 lower and upper confidence limits for our computed
parameters resulting from the fits (e.g. total infrared luminosity etc.) are calculated according to the
prescribed $\chi^2$ confidence intervals for one interesting
parameter, namely $\chi^2_{\rm min}+1$, where $\chi^2_{\rm min}$ is
the minimum $\chi^2$. 

\begin{figure}
\centering
\epsfig{file=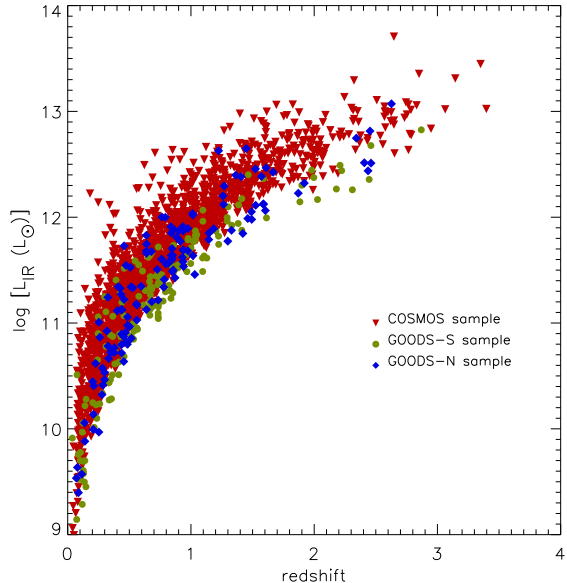, width=0.99\linewidth}
\caption{Total infrared luminosity as a function of redshift for the initial
 \textit{Herschel} sample of 2313 sources used for SED measurements (section \ref{sec:measurements}). }
\label{fig:LIR_z}
\end{figure}

Note that because our photometry only sparsely samples
the full IR SED, the parameters that characterise the best fit
SEDs within $\chi^2_{\rm min}+1$ are often degenerate and not well
constrained. In addition, each object in the sample is fit with the entire SK07 library
irrespective of the inherent luminosity of the templates; the total
infrared luminosity, $L_{\rm IR}$ (L$_{\odot}$), of each source is
computed by integrating the best matched SED model between 8 and
1000$\mu$m (see Fig. \ref{fig:LIR_z} for a plot of $L_{\rm IR}$ as a function of redshift). This allows us to stay clear of any
assumptions which link the SED shape to the
luminosity. However, it also implies that we cannot directly use some
of the SK07 parameters to describe our
sample, as they require scaling. One such parameter is the starburst size, as its impact on the SED
shape depends on the total input luminosity. Finally, the SK07
grid, although more flexible than most stand-alone SED libraries
currently in the public domain, is still too coarse to allow complete characterisation of the
physical properties of the sample. For these reasons, we opt to use one parameter to
describe the overall shape of the SK07 SED templates; we refer to this as the
flux ($\mathcal{F}$), calculated as log\,[$L_{\rm tot}$/4$\pi R^2$] in units of L$_{\odot}$\,kpc$^{-2}$, where
$L_{\rm tot}$ is the given luminosity of the template (not our
computed $L_{\rm IR}$) and $R$ is the
starburst size that corresponds to that template. In the SK07
formulation, for constant $A_{\rm V}$, as $R$ becomes larger, the dust mass increases as a
function of $R^2$ and hence becomes cooler, with the temperature being
a function of $L_{\rm tot}$/$R^2$. Hence, the larger $\mathcal{F}$ is, the more flux reaches the edge of the
starburst region and therefore the dust emission is
warmer. One can interpret high and low values of $\mathcal{F}$
as representative of systems with warmer/more compact and cooler/more extended
dust-emission respectively. 

In order to calculate dust temperatures, we use a modified black-body
function (a grey-body), of the form $B_{\lambda}(T)(1-e^{- \tau _{\lambda}})$, with a
wavelength dependent optical depth $\tau_{\lambda}=\tau_{100 \mu \rm m}(100 \mu \rm m/\lambda)^{\beta}$ (e.g. see
Klaas et al. 2001\nocite{Klaas01}) and a dust emissivity index $\beta$. We assume a low opacity limit, so approximate the
term $(1-e^{- \tau _{\lambda}})$ by $\lambda^{-\beta}$. Typical reported values of $\beta$ range between
1.5--2 (e.g. Dunne et al. 2000\nocite{Dunne00}, Lisenfeld, Isaak $\&$
Hills 2000\nocite{LIH00}) and we adopt $\beta$=1.5, consistent with studies of the far-IR emissivity of large
grains (Desert, Boulanger $\&$ Puget 1990\nocite{DBP90}). The
temperatures are derived by fitting all photometry at
$\lambda$\,$\ge$\,$\lambda_{ i_{\rm max}-1}$, where \textit{i} denotes
a \textit{Herschel} band (100, 160, 250, 350, 500\,$\mu$m) and \textit{i}$_{\rm
  max}$ is the band which corresponds to the maximum flux ($\nu
f_{\nu}$); the 24\,$\mu$m photometry is never included in
the fitting. 
Note that although some studies have shown that a two-temperature
greybody model (e.g. Klaas et al. 2001\nocite{Klaas01}; Dunne $\&$
Eales 2001\nocite{DE01}), is a more accurate description of far-IR
dust-emission, this would not work with our available
photometry; much longer wavelength data would be needed especially
for sources at high redshift. However, in any case, our aim is to measure the
average dust temperature of the far-IR peak for each source,
thus a single temperature greybody model is required; our
method gives a temperature which is most representative of the peak dust emission.

\begin{figure}
\centering
\epsfig{file=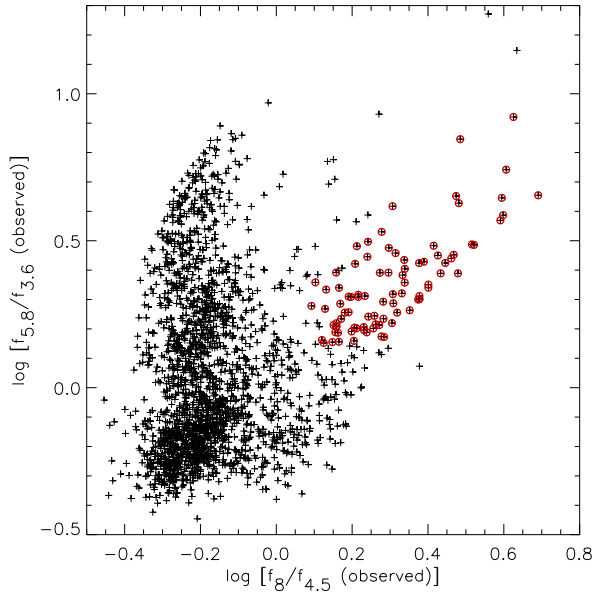, width=0.99\linewidth}
\caption{Observed \textit{Spitzer}/IRAC colours ($f_{5.8}$/$f_{3.6}$
  versus $f_{8}$/$f_{4.5}$ ) for the \textit{Herschel} sample (black
  crosses). Red circles denote sources which are identified as
  AGN-dominated in the near/mid-IR using the IRAC criteria outlined in Donley et al. (2012).}
\label{fig:AGN}
\end{figure}

\section{Dealing with AGN contamination}
\label{sec:AGN}

As this work targets the properties of the star-forming
galaxy population, objects whose infrared energy budget potentially includes a significant contribution from an
AGN need to be removed from the sample. Although the fraction of \textit{Herschel} galaxies found to host AGN
is high ($\sim$30 per cent; Symeonidis et
al. 2011b\nocite{Symeonidis11b}), it has been shown that AGN in
far-IR-selected galaxies do not often dominate the infrared or total energy
budget of the system. According to an energy balance argument, if the AGN
is energetic enough to contribute significantly to the infrared emission
of a starburst galaxy, then its signature is likely to emerge in the mid-IR
part of the SED in the form of a power-law continuum (e.g. Symeonidis et al. 2010\nocite{Symeonidis10}). As a result, the
most suitable way to identify such objects is by examining their colours in the
\textit{Spitzer}/IRAC (3.6, 4.5, 5.8, 8\,$\mu$m) bands. Until recently, the most commonly
used IRAC AGN selection criteria have been those presented in Lacy et al. (2004\nocite{Lacy04}) and
Stern et al. (2005\nocite{Stern05}). However, as shown in Yun et
al. (2008\nocite{Yun08}) and Donley et al. (2012\nocite{Donley12}),
there is a non-negligible chance that IR/submm selected galaxies will
be erroneously identified as AGN-dominated in the IRAC bands. In
addition, Hatziminaoglou et al. (2009\nocite{Hatziminaoglou09}) reported
`cross-talk' between the AGN and SB loci in the Lacy et al. (2004) diagram.
Here, we use the Donley et al. (2012) IRAC criteria, shown to be effective in picking out AGN and
sufficiently robust against
misidentifications. Fig. \ref{fig:AGN} shows $f_{5.8}$/$f_{3.6}$ colour
against $f_{8}$/$f_{4.5}$ colour for the \textit{Herschel}
sample. Indicated in red are the sources which satisfy the Donley et
al. (2012) criteria and are hence classified as AGN-dominated in the
near/mid-IR. These 87 (out of 2313) objects, $\sim$\,4 per
cent, are subsequently excluded from the sample, leaving 2226
sources. Note that this is not the fraction of AGN hosted by
\textit{Herschel} sources, rather it is the
fraction of objects where the AGN could contribute substantially in
the mid-infrared and hence interfere with our analysis. A 4 per cent
fraction of sources with AGN-dominated near/mid-IR SEDs
is in line with results from Symeonidis et al. (2010) and Pozzi et
al. (2012\nocite{Pozzi12}) who show that AGN rarely contribute more
than 20 per cent in the IR emission of far-IR detected systems (see also
Hatziminaoglou et al. 2010; Page et al. 2012\nocite{Page12}; Rosario et al. 2012\nocite{Rosario12};
Nordon et al. 2012\nocite{Nordon12}).

\begin{figure}
\centering
\begin{tabular}{c}
\epsfig{file=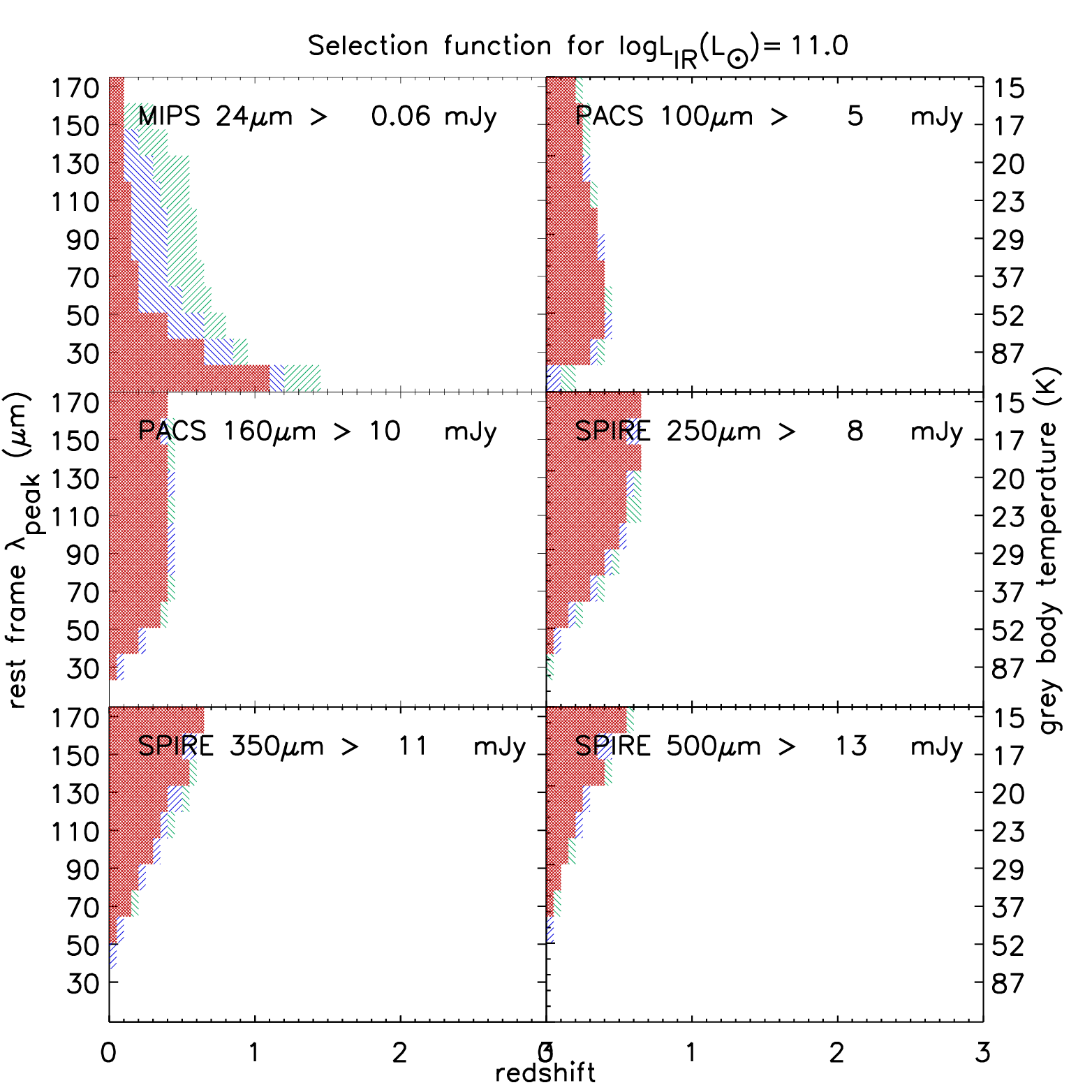,width=0.97\linewidth,clip=} \\
\epsfig{file=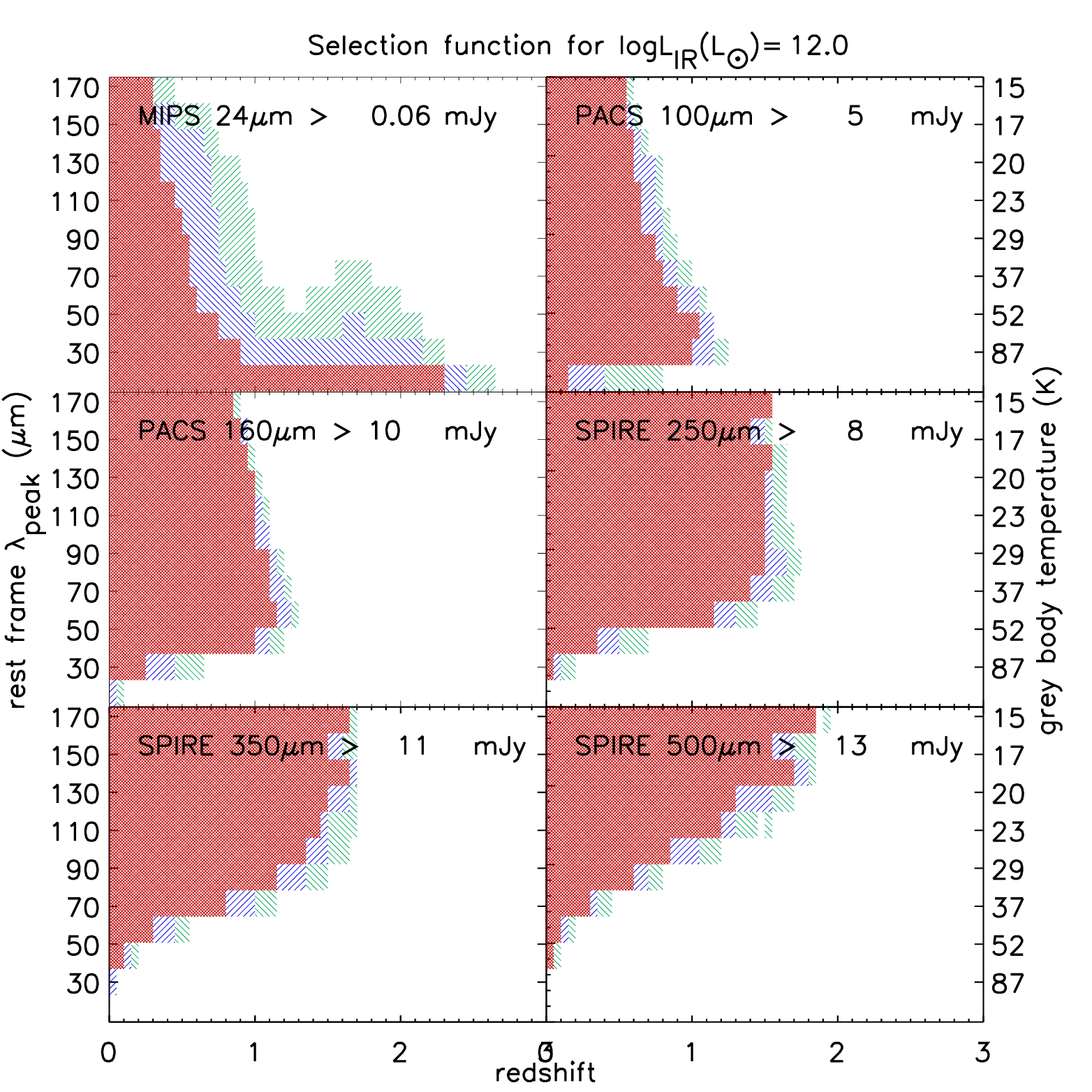,width=0.97\linewidth,clip=} \\
\end{tabular}
\caption{log\,($L_{\rm
    IR}$/L$_{\odot}$)=11 and log\,($L_{\rm IR}$/L$_{\odot}$)=12
  selection functions for PACS and SPIRE, constructed using the SK07 model library; see also
  Symeonidis et al. (2011a). The plot shows SED peak wavelength (left y-axis) and grey body
  temperature (right y-axis) as a function of redshift. At any
  redshift slice and for a given flux density limit, the red region indicates that all SED shapes of the
  corresponding peak wavelength are detectable, whereas the blue
and green dashed patterns indicate regions where only 90 and 70 per
cent of the SK07 templates are recovered respectively. In the unshaded areas less than 70 per cent of
templates are detectable, reaching zero above a certain
redshift. The flux density limits used to construct these diagrams are
0.06, 5, 10, 8, 11 and 13 mJy for the 24, 100, 160, 250, 350 and
500\,$\mu$m bands respectively,  corresponding to the COSMOS 3\,$\sigma$ flux density
limits.}
\label{fig:selection}
\end{figure}

\section{Dealing with selection effects}
\label{sec:selection}

\subsection{\textit{Herschel} selection}
\label{sec:herschel_selection}
To accurately characterise the aggregate properties of the IR-luminous
population and their
evolution with redshift, there should be no bias with regard to the SED types we can
observe, particularly regarding the far-IR where all our measurements
are performed. As a result, the accuracy of our work rests on
minimising selection biases and assembling a sample within an
unbiased part of the $L-z$ parameter space. 
For this purpose, we use the method described in Symeonidis et
al. (2011a) to examine the selection functions of the PACS and SPIRE
bands, mapping out an SED-redshift-luminosity parameter space at the
flux density limits of the GOODS and COSMOS surveys used in this work.

Fig. \ref{fig:selection},
shows the $L_{\rm
  IR}$\,$\sim$10$^{11}$\,L$_{\odot}$ and 10$^{12}$\,L$_{\odot}$ selection functions for MIPS 24\,$\mu$m,
PACS 100, 160\,$\mu$m and SPIRE 250, 350 and 500\,$\mu$m at the COSMOS flux density limits. As also explained in detail in Symeonidis et al. (2011a), Fig. \ref{fig:selection} is created by using all
SED templates from the SK07 library, normalising them to the required total
IR luminosity and then scaling and redshifting them to the observed
frame. Each SED template is then convolved with the MIPS, PACS and
SPIRE filter transmission curves in
order to get the weighted integrated flux within each filter. We
subsequently perform a colour correction (according to the
prescription in
the instruments' observer manuals) in order to obtain a monochromatic
flux density in each band, derived with the same spectral shape used
to calculate the measured flux density of real sources. 
For each band we then compare our template monochromatic flux density to
the flux density limit, in order to determine whether
an object with the given redshift, luminosity and SED shape would be
part of our sample. This results in the selection functions presented in
Fig. \ref{fig:selection}. Red thick patterns mark the regions where all templates of a
given peak wavelength are detected, whereas the blue
and green dashed patterns indicate regions where only 90 and 70 per
cent of the SK07 templates are recovered respectively. The detection
rate relates to variations in SED shape; for example, for the MIPS
24\,$\mu$m selection function, only luminous SEDs with strong PAH features will
be detected at $z\sim$1.7. In the unshaded areas less than 70 per cent of
templates are detectable, a fraction which reduces to zero above a
certain redshift.

\begin{figure}
\centering
\begin{tabular}{c}
\epsfig{file=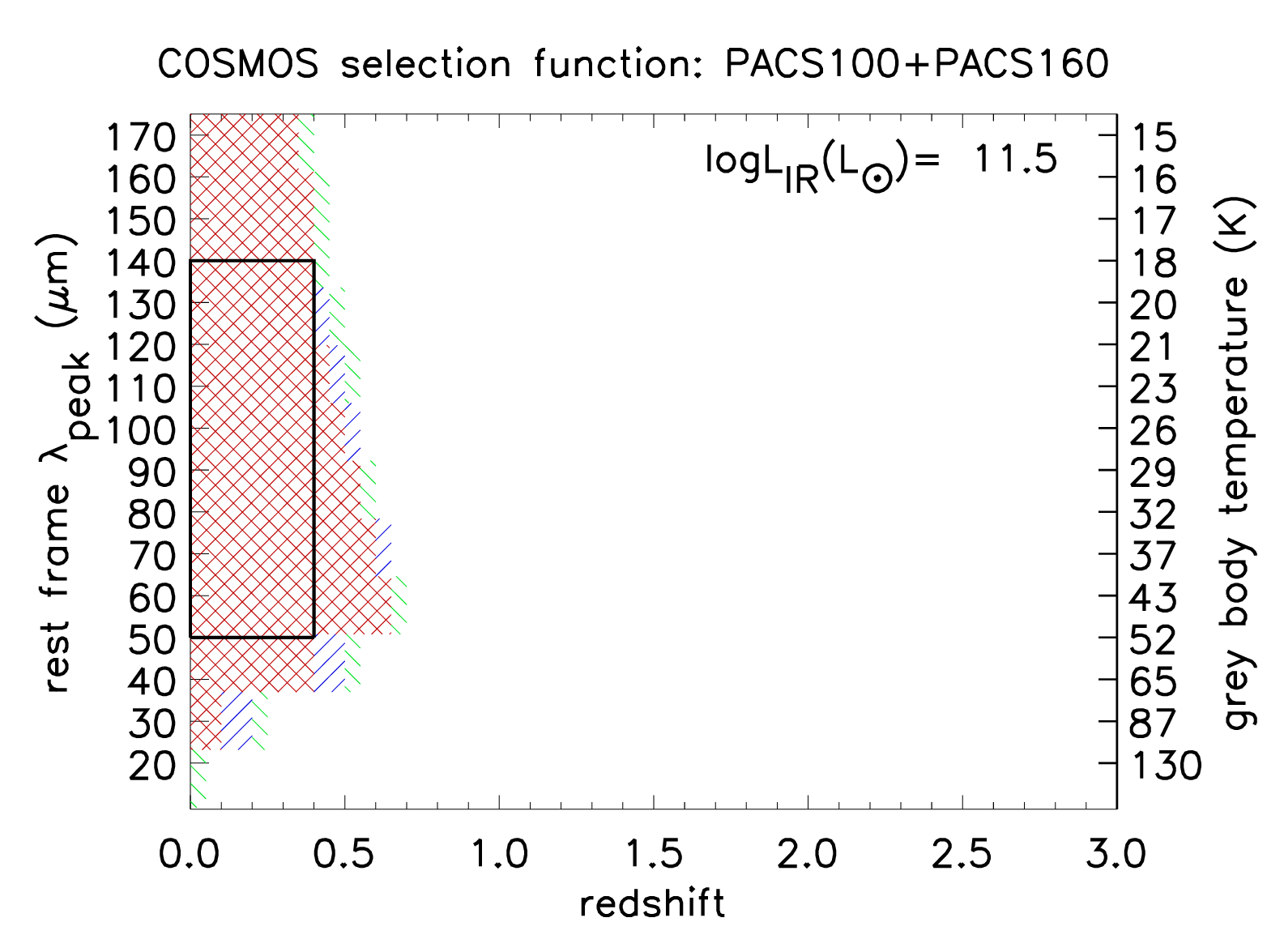,width=0.88\linewidth,clip=}\\
\epsfig{file=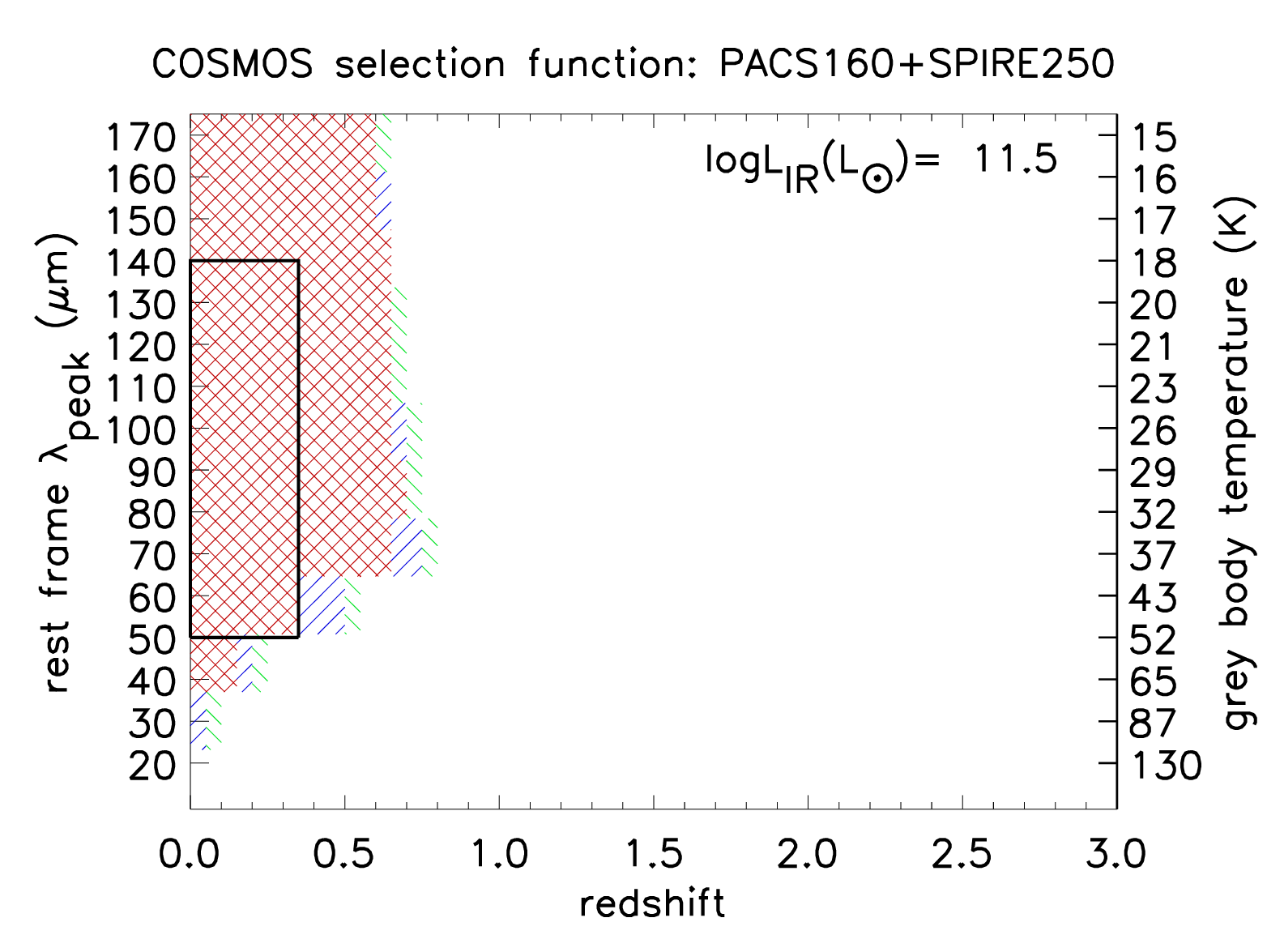,width=0.88\linewidth,clip=}\\
\epsfig{file=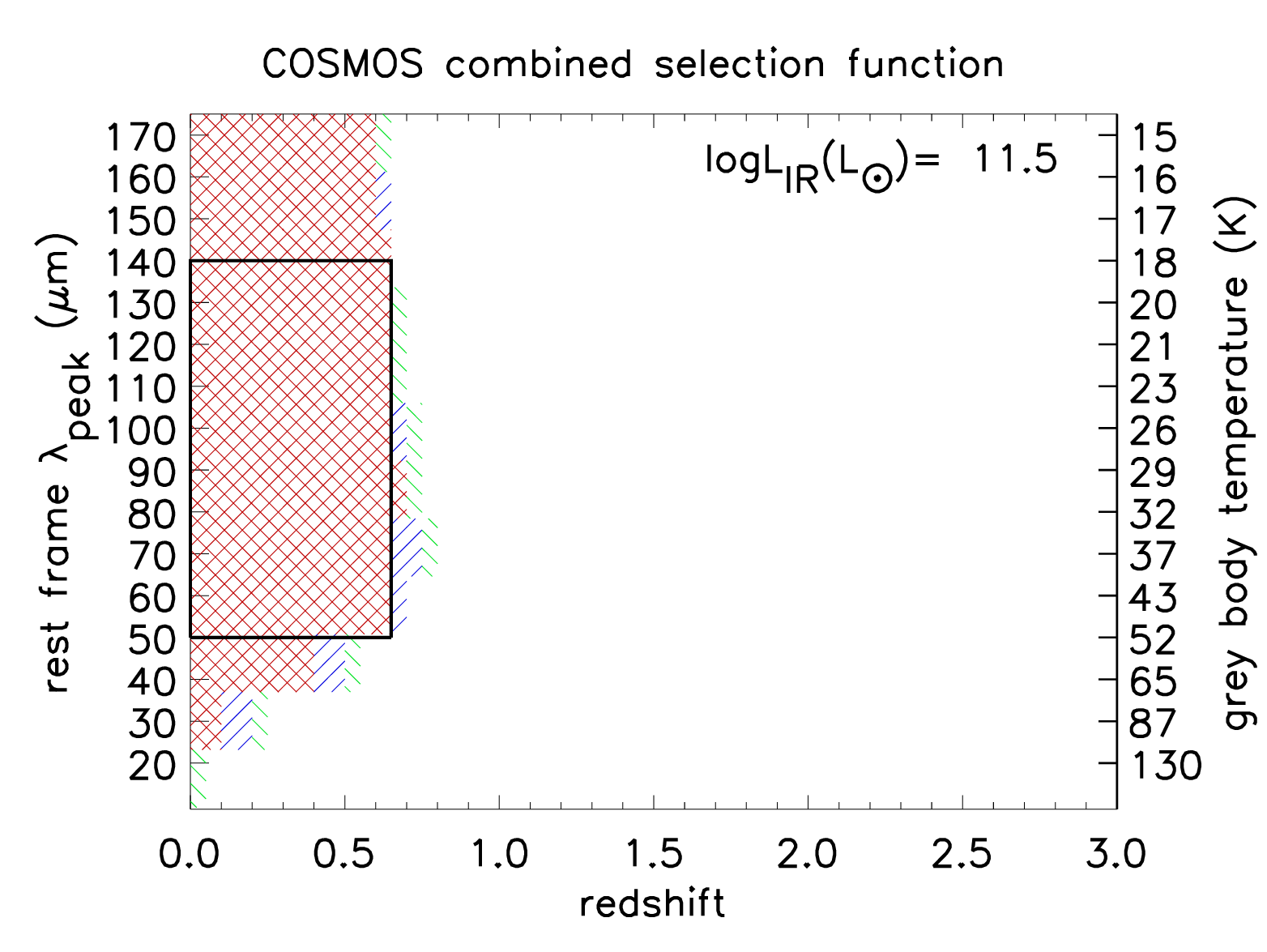,width=0.88\linewidth,clip=}\\
\end{tabular}
\caption{Selection functions for log\,($L_{\rm
    IR}$/L$_{\odot}$)=11.5 using the PACS 100+160\,$\mu$m criterion (top
  panel), the PACS 160 + SPIRE 250\,$\mu$m criterion (middle panel)
  and the two used in disjunction ([100+160\,$\mu$m] OR [160+250\,$\mu$m]). The plot shows SED peak wavelength (left y-axis) and grey body
  temperature (right y-axis) as a function of redshift. At any
  redshift slice and for a given flux density limit, the red region indicates that all SED shapes of the
  corresponding peak wavelength are detectable, whereas the blue
and green dashed patterns indicate regions where only 90 and 70 per
cent of the SK07 templates are recovered respectively. In the unshaded areas less than 70 per cent of
templates are detectable, reaching zero above a certain
redshift. The flux density limits used to construct these diagrams are
5, 10 and 8 mJy for the 100, 160 and 250 $\mu$m bands respectively,
corresponding to the 3$\sigma$ flux density limits in
  COSMOS. The black box in each panel outlines the redshift range
  where all SED shapes with peak wavelengths 50-140\,$\mu$m are
  detectable --- note that the redshift range is larger in the 3rd panel.}
\label{fig:selection_combined}
\end{figure}

\begin{figure}
\centering
\begin{tabular}{c}
\epsfig{file=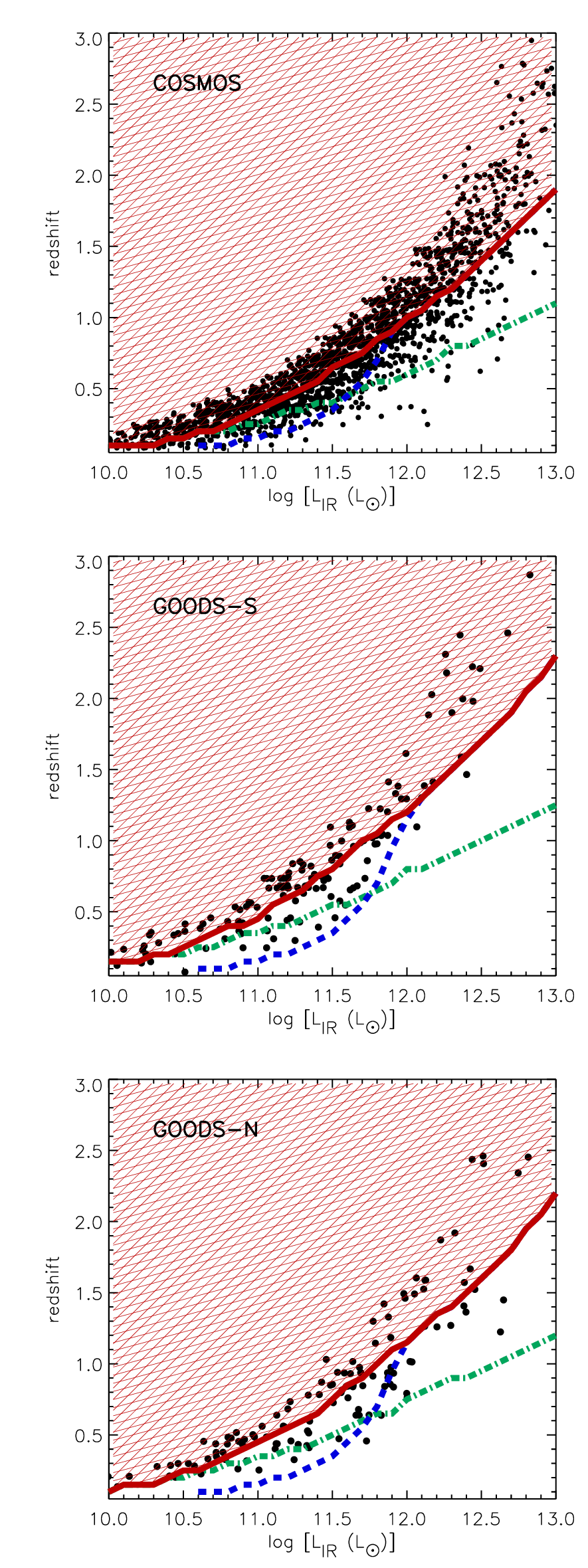,width=0.9\linewidth,clip=}\\
\end{tabular}
\caption{Redshift versus $L_{\rm IR}$ for the \textit{Herschel}
  sources (black points) in COSMOS, GOODS-S, GOODS-N. The regions below the green
  and blue dashed lines mark the complete parameter space derived from the [100+160\,$\mu$m]
  and [160+250\,$\mu$m] selection functions respectively. The red curve corresponds to the
  parameter space traced out by the two criteria in disjunction ([100+160\,$\mu$m] OR
  [160+250\,$\mu$m]), with the hatched red pattern indicating the region
  of incompleteness. In the work presented in this paper we use only sources which lie
  below the red curve, which includes any source with $50
  <\lambda_{\rm peak} (\rm \mu m) < 140$, or, equivalently, temperature
within $18 \lesssim T (\rm K) \lesssim 52$. }
\label{fig:selection_sample}
\end{figure}

Note that the selection functions of PACS and SPIRE 
overlap significantly, suggesting that the SEDs of most sources
will be fully sampled by \textit{Herschel} photometry. However, there
are large differences between the MIPS/24\,$\mu$m and the
\textit{Herschel} selection functions, especially at 500\,$\mu$m. This is not surprising as they probe
different parts of the SED and it is expected that the long
wavelength one eventually turns to favour very cold sources and the
short wavelength one eventually turns to favour warm
sources. The 24$\mu$m selection is a steep function of SED shape: SEDs with a
significant warm dust component are favoured up to very high
redshifts. The small area covered by the red pattern implies that it
is only over a small redshift range that all SED shapes are
recoverable. However, the large blue and green shaded regions indicate that
cold SEDs \emph{can} be detected up to high redshift, as long as they
do not have a high far-IR to mid-IR ratio.

Ignoring the 24\,$\mu$m selection for the moment, Fig.
\ref{fig:selection} shows that our far-IR selection criteria of a
detection at [100+160\,$\mu$m] OR [160+250\,$\mu$m] will
result in the largest number of sources detected in an unbiased part of $L$-$z$
space. This is more clear in Fig. \ref{fig:selection_combined} which shows
the [100+160$\mu$m] selection function (top panel), the
[160+250\,$\mu$m] selection function (middle panel) and the two
criteria in disjunction ([100+160\,$\mu$m] OR [160+250\,$\mu$m]; lower
panel) for log\,($L_{\rm IR}$/L$_{\odot}$)=11.5, at the COSMOS limits. The black boxes outline
the extent in redshift whereby all SEDs with peak between 50-140\,$\mu$m
will be detected. This translates to a temperature range of 18--52\,K,
using the Wien displacement law for a $\nu f_{\nu}$ grey body, $T(K)\sim\frac{hc}{(4+\beta) k \lambda_{\rm peak}}$,
where h is the Planck constant, c is the speed of light in a vacuum
and k is the Boltzmann constant and we take the dust emissivity
($\beta$) to be 1.5.
This choice of peak wavelength/temperature range is the best compromise between the range of SED types
probed and the number of objects studied. As we shall see in section
\ref{sec:results} increasing that range would not have changed the
measured average properties of the sample but would have significantly
reduced the statistics. 
The black box in the lower panel of Fig. \ref{fig:selection_combined} shows that using the two criteria
in disjunction, i.e. [100+160\,$\mu$m] OR [160+250\,$\mu$m], allows
log\,($L_{\rm IR}$/L$_{\odot}$)=11.5 sources to be selected up to much higher
redshifts, than when these criteria are used separately.

\begin{figure}
\centering
\epsfig{file=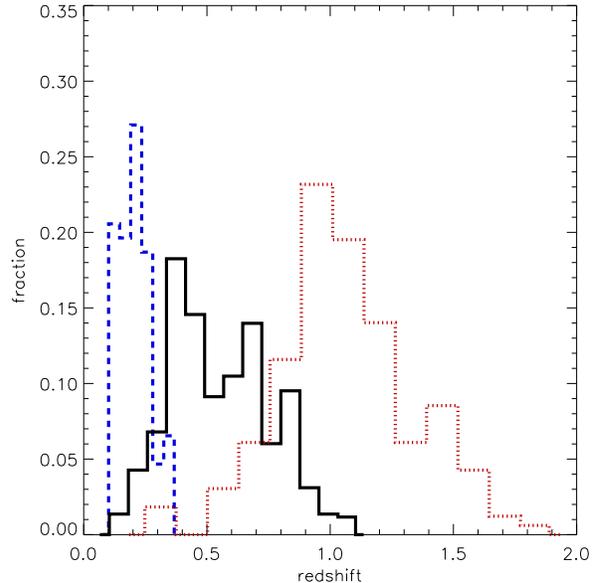, width=0.99\linewidth}
\caption{The redshift distribution of the final \textit{Herschel}
  sample of 1159 sources used in this work; dashed blue line for normal
  IR galaxies (NIRGs), solid black line for LIRGs and red
  dotted line for ULIRGs. }
\label{fig:zdistribution}
\end{figure}

With the aid of the combined selection functions, we now define the complete $L_{\rm IR}$-$z$
parameter space for each survey (GOODS-N $\&$ S, COSMOS), shown in
Fig. \ref{fig:selection_sample}, where the curves
separate the complete (below curve) and incomplete (above curve) part
of the parameter space. The space below the curves indicates the
$L_{\rm IR}$-$z$ range, where any source with SED peak wavelength within
$50<\lambda_{\rm peak} (\rm \mu m)<140$, or, equivalently, temperature
within $18 \lesssim T (\rm K) \lesssim 52$, is detectable. 
As mentioned earlier, the [100+160\,$\mu$m] criterion (green curve in
Fig. \ref{fig:selection_sample}), picks up more sources at low
redshift but excludes more high redshift sources because its
selection function quickly turns over to favour warm SEDs providing a
limited unbiased $L$-$z$ space at high redshift. On the
other hand the [160+250\,$\mu$m] criterion (blue curve in
Fig. \ref{fig:selection_sample}) performs poorly at low redshift,
because the SPIRE 250\,$\mu$m which mainly drives the combined
selection function, largely favours cold sources. At $z$\,$\sim$1, the
combination of 160 and 250\,$\mu$m turns over as now these bands are sampling 80 and 125$\mu$m respectively,
covering the bulk of IR emission. The combined
criterion of [100+160\,$\mu$m] OR [160+250\,$\mu$m] (red curve in
Fig. \ref{fig:selection_sample}) is what we thus use to select the
final sample used in this work. Note that it is the 160\,$\mu$m
band that principally drives the combined selection function, whereas the PACS
100\,$\mu$m and SPIRE 250\,$\mu$m in essence provide an additional
band in the infrared, vital for our analysis. However, as they complement each
other well, such that most SEDs missed at 100\,$\mu$m are picked up at 250\,$\mu$m and vice
versa, the complete region below the red curve in Fig. \ref{fig:selection_sample}
includes many sources which are in the incompleteness regions of both the blue
and green curves. In addition, this selection criterion ensures that
the SED peak is well sampled for most sources up to $z\sim2$. 

Although the selection outlined above is quite conservative since it
is unlikely that all SK07 models with 50$< \lambda_{\rm peak} (\rm \mu m)
<$140 are representative of real objects, it nevertheless allows us to
perform our analysis within a bias-free framework with respect to the
PACS and SPIRE surveys. Hereafter, our study concerns only the 1159 sources within the
complete parameter space below the red curves in
Fig. \ref{fig:selection_sample}. The redshift distribution of the
final \textit{Herschel} sample is shown in Fig. \ref{fig:zdistribution}.

\begin{figure}
\centering
\epsfig{file=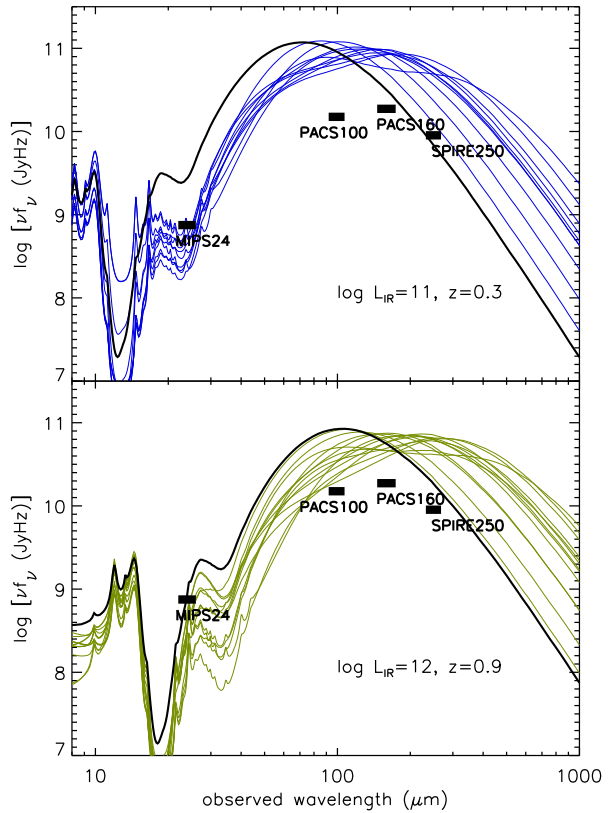,width=0.99\linewidth,clip=}\\
\caption{Example SED templates from the SK07 library, shown in the
  observed frame. In the top panel
  they are normalised to a luminosity of $L_{\rm
    IR}$=10$^{11}$\,L$_{\odot}$ and then scaled and
redshifted to $z$=0.3. In the lower panel they are normalised to a
luminosity of $L_{\rm IR}$=10$^{12}$\,L$_{\odot}$ and scaled and redshifted to
$z$=0.9. At these luminosities and redshifts, these
templates would not be detected down to the 60\,$\mu$Jy COSMOS 24\,$\mu$m 
flux density limit. However they are detected in the \textit{Herschel}
bands at the COSMOS 3\,$\sigma$ 100, 160, 250\,$\mu$m limits, also
shown in both panels. For comparison, the SED of Arp220 (black SED; taken from SK07) is also plotted normalised at
the appropriate luminosity and redshifted. Note that at $z$=0.9, it is
only just detected. }
\label{fig:seds_det}
\end{figure}

\begin{figure}
\centering
\epsfig{file=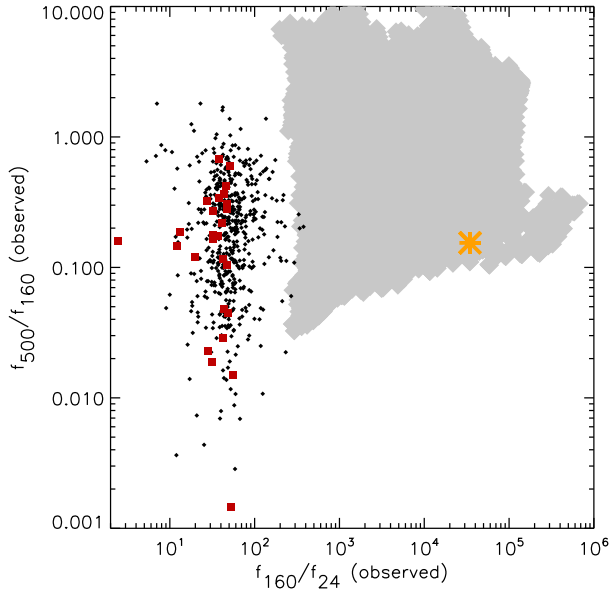,width=0.99\linewidth,clip=}\\
\caption{ Observed $f_{500}$/$f_{160}$ versus
  $f_{160}$/$f_{24}$ colour. The grey shaded region shows the colours
  of SK07 templates, detected in the \textit{Herschel} bands within the complete
  COSMOS $L-z$ parameter space outlined in Fig. \ref{fig:selection_sample}, but missed by the
  COSMOS 24\,$\mu$m selection (some examples of these templates are shown in
  Fig. \ref{fig:seds_det}). The black dots are the COSMOS
  sample and red squares are the GOODS-S and GOODS-N sources within
  the COSMOS $L-z$ completeness region (Fig. \ref{fig:selection_sample}). The orange asterisk are the
colours for the SK07 SED of Arp220 redshifted to $z=$1.5.}
\label{fig:ratio160_24}
\end{figure}

\subsection{The 24\,$\mu$m selection}
\label{sec:24um_selection}
In section \ref{sec:herschel_selection} we assembled a complete sample
with respect to the \textit{Herschel} bands, which cover the part of the SED
primarily used in our study (see section \ref{sec:results}). It is now
important to examine whether the requirement for a 24\,$\mu$m detection affects the
completeness of this sample. 

Fig. \ref{fig:selection} shows that although the 24\,$\mu$m and
160\,$\mu$m selection functions cover approximately the same redshift
range, some SEDs are systematically missed at 24\,$\mu$m. We
investigate this further by aiming to answer the following questions:
what types of SEDs are missed, how common are sources with such SED types and
how does this affect our results.
The first question is easier to answer. These SEDs are ones with
high far-to-mid-IR ratio, resulting from a combination
of parameters in the SK07 formulation, such as high extinction
($A_{\rm V}$$>$70) and/or low luminosity and/or low dust
density within the hot spot, with a detection rate that is also
redshift dependent. 
Examples of these SEDs (in the observed frame) are presented in
Fig. \ref{fig:seds_det}, and Fig. \ref{fig:ratio160_24} shows
the $f_{500}$/$f_{160}$ -- $f_{160}$/$f_{24}$ colour space these cover; in both figures we assume the COSMOS
flux limits. 
Fig. \ref{fig:seds_det} shows that such templates are characterised
by steep mid-IR continua and strong silicate
absorption features at 9.7 and 18\,$\mu$m, a result of high extinction
in the SK07 formulation. Note that these
SED types are easily detected in the \textit{Herschel} bands. 
In both panels of Fig. \ref{fig:ratio160_24}, we also show the SED of
Arp220 (taken from SK07), normalised at the appropriate luminosity and
redshift. Arp220 is one of the most optically thick
ULIRGs known (e.g. Papadopoulos et al. 2010\nocite{Papadopoulos10}), so it is interesting to examine whether it would be
detected in our sample. The top panel shows that at low redshift it
would be detected in all bands, however, at $z$=0.9 it is
only just detected at 24\,$\mu$m, but easily detected with
\textit{Herschel}. As this SED is redshifted further, it will be
missed by the 24\,$\mu$m survey, suggesting that optically thick SEDs,
with deep silicate absorption, would not be in our sample at high
redshift ($z\gtrsim$1).

\begin{figure}
\centering
\epsfig{file=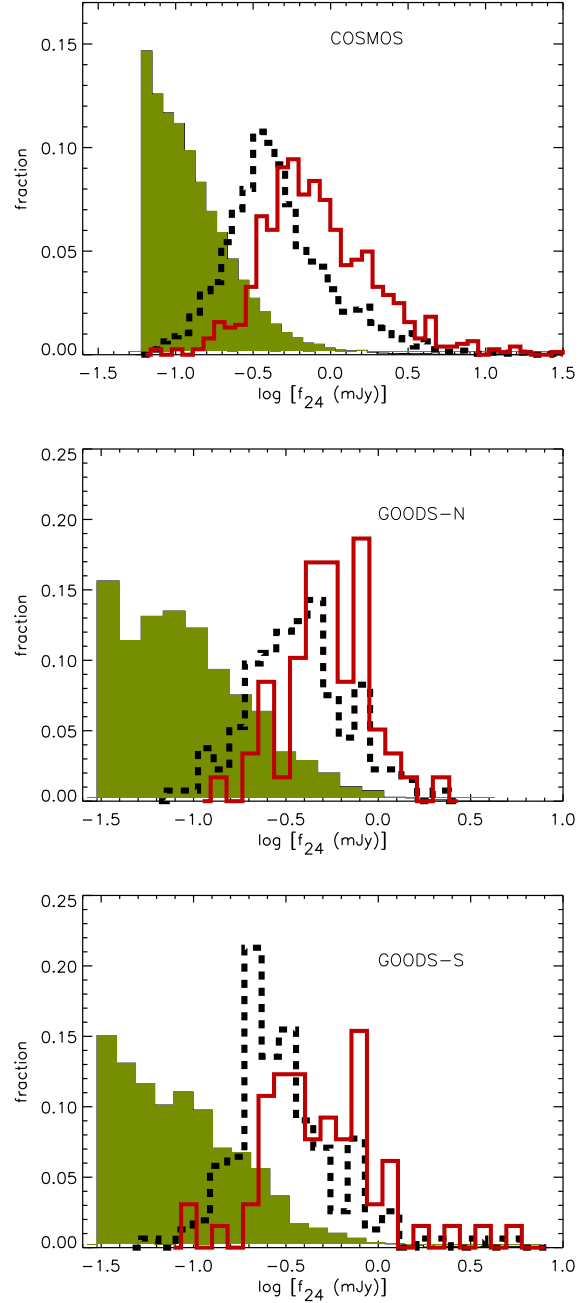,width=0.96\linewidth,clip=}\\
\caption{ The 24\,$\mu$m flux density distribution of the 24\,$\mu$m
  population in COSMOS, GOODS-N and GOODS-S (green filled-in histogram) compared to
  that of the final
  \textit{Herschel} sample before (black dashed histogram) and after the
\textit{Herschel} completeness criteria are applied (red solid
histogram; see section \ref{sec:herschel_selection}). }
\label{fig:selectiontest}
\end{figure}

Fig. \ref{fig:ratio160_24} shows the observed $f_{500}$/$f_{160}$ versus $f_{160}$/$f_{24}$ colours for the
COSMOS sample as well as for sources in the GOODS samples which
fall within the COSMOS completeness region shown in Fig. \ref{fig:selection_sample}. The grey-shaded
region is the parameter space occupied by the SED templates which are
detected in the \textit{Herschel} bands within the complete $L_{\rm IR}-z$ 
parameter space for COSMOS outlined in Fig. \ref{fig:selection_sample}, but are missed by
the 24\,$\mu$m selection. 
Some overlap between the samples and shaded region should be
expected, as some sources could have high $f_{160}$/$f_{24}$
ratios because of high $f_{160}$ rather than a low $f_{24}$. However,
we see very little overlap, suggesting that the detected and
non-detected SEDs cover a significantly different part of
parameter space in terms of their far-to-mid-IR colour. 
This implies that, on average, SEDs are missed by the 24\,$\mu$m selection because
of a particular feature that reduces the amount of
24\,$\mu$m observed flux, such as a silicate feature, rather than an
inherently steep far-to-mid-IR continuum. This also becomes obvious from the values of $f_{160}$/$f_{24}$
ratios that some templates have: such high values of
$f_{160}$/$f_{24}$, up to 6 orders of magnitude, can only be
caused by a strong 9.7\,$\mu$m silicate feature; c.f. with what is
observed for the Arp\,220 SED redshifted at
$z=$1.5, where the 9.7\,$\mu$m feature falls in the 24\,$\mu$m band. 
In terms of the $f_{500}$/$f_{160}$ ratio, there are only a handful of
\textit{Herschel} sources with $f_{500}$/$f_{160}$\,$>$1, whereas a
significant fraction of the templates in the grey region have such
cold colours. This is not surprising as we would expect some of the
templates that are missed at 24\,$\mu$m to be overall colder and hence have higher $f_{500}$/$f_{160}$; see some
examples in Fig. \ref{fig:seds_det}. 

To answer the second question `how common are these SED types?', we
compare the GOODS (N and S) to the COSMOS
colours in Fig. \ref{fig:ratio160_24}. As mentioned earlier, the GOODS
sources shown in Fig. \ref{fig:ratio160_24} are within the COSMOS
completeness region mapped out in Fig. \ref{fig:selection_sample}. However, for GOODS,
the 24\,$\mu$m flux density limit is twice as deep as it is in COSMOS,
so in principle these GOODS sources could have SEDs with
f$_{160}$/f$_{24}$\,$>$330. We do not find any such sources, in fact
we note that GOODS objects have f$_{160}$/f$_{24}$ ratios within the
range covered by COSMOS, suggesting that SEDs with high f$_{160}$/f$_{24}$ (up to
$z$$\sim$2) are rare.

In Fig. \ref{fig:selectiontest} we examine the flux
density distribution of the 24\,$\mu$m population in the 3 fields under study, in comparison to the distribution of
\textit{Herschel} sources before and after the
\textit{Herschel} completeness criteria are applied (section
\ref{sec:herschel_selection}). Note the significant offset between the
distributions of \textit{Herschel} sources and that of the 24\,$\mu$m population, suggesting that 
\textit{Herschel} flux densities are intrinsically correlated with bright 24\,$\mu$m
flux densities. This is unlikely to be an artifact of our selection,
as we are using only `isolated' 24\,$\mu$m objects, so there is no
reason why, in principle, a \textit{Herschel} source cannot be
associated with a faint 24\,$\mu$m source. It is immediately obvious
from Fig. \ref{fig:selectiontest} that the fraction of sources which
would not be part of our sample because of the requirement of a 24\,$\mu$m detection is very
small. 
Assuming normally distributed flux densities and by calculating the mean
and standard deviation of each distribution, we can compute $n \sigma$ where $n$ is the number
of standard deviations ($\sigma$), at the location of the 24\,$\mu$m
flux density limit. This gives a rough indication of the fraction of
sources that are likely missed due to the 24\,$\mu$m selection. 
Before the \textit{Herschel} completeness criteria are applied, we estimate that 0.1, 1.4 and 0.01
per cent of sources are unaccounted for in GOODS-S, COSMOS and GOODS-N respectively. After these criteria are applied, the fraction of missing sources in
COSMOS goes down to 0.2 per cent, whereas for the GOODS fields it is
less than 0.06 per cent. Although these are rough estimates and rest
on the assumption of normally distributed flux densities, they do
indicate that the fraction of sources missed by the 24\,$\mu$m selection
is very small once our final sample is assembled in the complete
$L-z$ parameter space using the  \textit{Herschel} selection functions
(section \ref{sec:herschel_selection}).

\begin{figure}
\centering
\epsfig{file=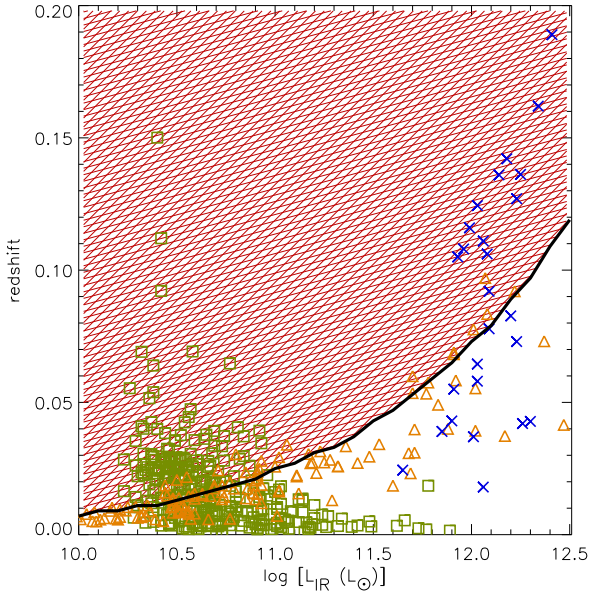, width=0.99\linewidth}
\caption{Plot of redshift versus total infrared luminosity for the
  local ($z\lesssim$0.1) samples (see section \ref{sec:localsample});
  blue crosses: Clements et
  al. (2010), orange triangles: Hwang et al. (2010) and green squares: Buat et al. (2010). The region below the black curve
  marks the complete parameter space corresponding to the \textit{IRAS} 1\,Jy 60\,$\mu$m selection
  function, with the hatched red pattern indicating the region
  of incompleteness. Only sources below the
  curve are used in our analysis.}
\label{fig:iras_selection}
\end{figure}

In light of this analysis, we conclude that (i)  the
SEDs missed by the 24\,$\mu$m selection have high far-to-mid-IR
ratios, mainly as a result of high optical
depth in the mid-IR and deep silicate features, (ii) these
SEDs are not a common
occurrence amongst IR-luminous galaxies up to $z \sim$2 and (iii) our final
\textit{Herschel} sample can be assumed to be complete in terms of the
SED types we can detect. 
Our findings are consistent with the study of Magdis et al. (2011) who
found the \textit{Herschel} 24\,$\mu$m dropouts to constitute a few per cent of the IR-luminous galaxy
population at high redshift and concluded that they
must be sources with stronger silicate absorption features --- see also
Roseboom et al. (2010); Lutz et al. (2011) for discussion of
24\,$\mu$m selection effects. 
Finally the third question of `how would
the fraction of sources missed affect our results' is discussed at the
end of section \ref{sec:evolution}.

\subsection{Local sample selection}
\label{sec:localsample}

In order to compare the \textit{Herschel} sample to
analogous sources in the nearby ($z\lesssim$0.1) Universe, we also assemble an
\textit{IRAS}-selected local sample of $L_{\rm
  IR}$\,$>$\,10$^{10}$\,L$_{\odot}$ galaxies by combining sources from
Clements et al. (2010\nocite{CDE10}), Hwang et
al. (2010\nocite{Hwang10}) and Buat et
al. (2010\nocite{Buat10}). For the Clements et al. (2010) objects, we retrieve their published single
greybody temperatures and total infrared luminosities calculated using \textit{IRAS} 60 and
100\,$\mu$m data as well as SCUBA data. For the other two samples
(Hwang et al. 2010 and Buat et al. 2010), we use their computed
total infrared luminosities and
calculate greybody temperatures by fitting a greybody function
of emissivity $\beta$=1.5 (see section \ref{sec:measurements}) to the \textit{IRAS} and
\textit{AKARI} fluxes at $\lambda$\,$\ge$60\,$\mu$m. In all cases, the SEDs of the local sources
have some coverage at $\ge$100\,$\mu$m, either because of \textit{AKARI}
or SCUBA data. 

In order to combine these samples for subsequent analysis,
we select all sources which are in the complete $L-z$ region
of the \textit{IRAS}/60\,$\mu$m selection function down to a flux
density of 1\,Jy (see method in section \ref{sec:selection} and Symeonidis et al. 2011a for the \textit{IRAS}
selection function). Fig. \ref{fig:selection_sample} shows the
curve dividing complete and incomplete parts of parameter
space. We cut the local samples to include only sources in the
complete parameter space, where any
source with $f_{60}>$1\,Jy and SED peak wavelength within
$50<\lambda_{\rm peak} (\rm \mu m)<140$, or, equivalently, temperature
within $18 \lesssim T (\rm K) \lesssim 52$, is detectable.

\section{Results}
\label{sec:results}

\begin{figure*}
\centering
\begin{tabular}{c}
\epsfig{file=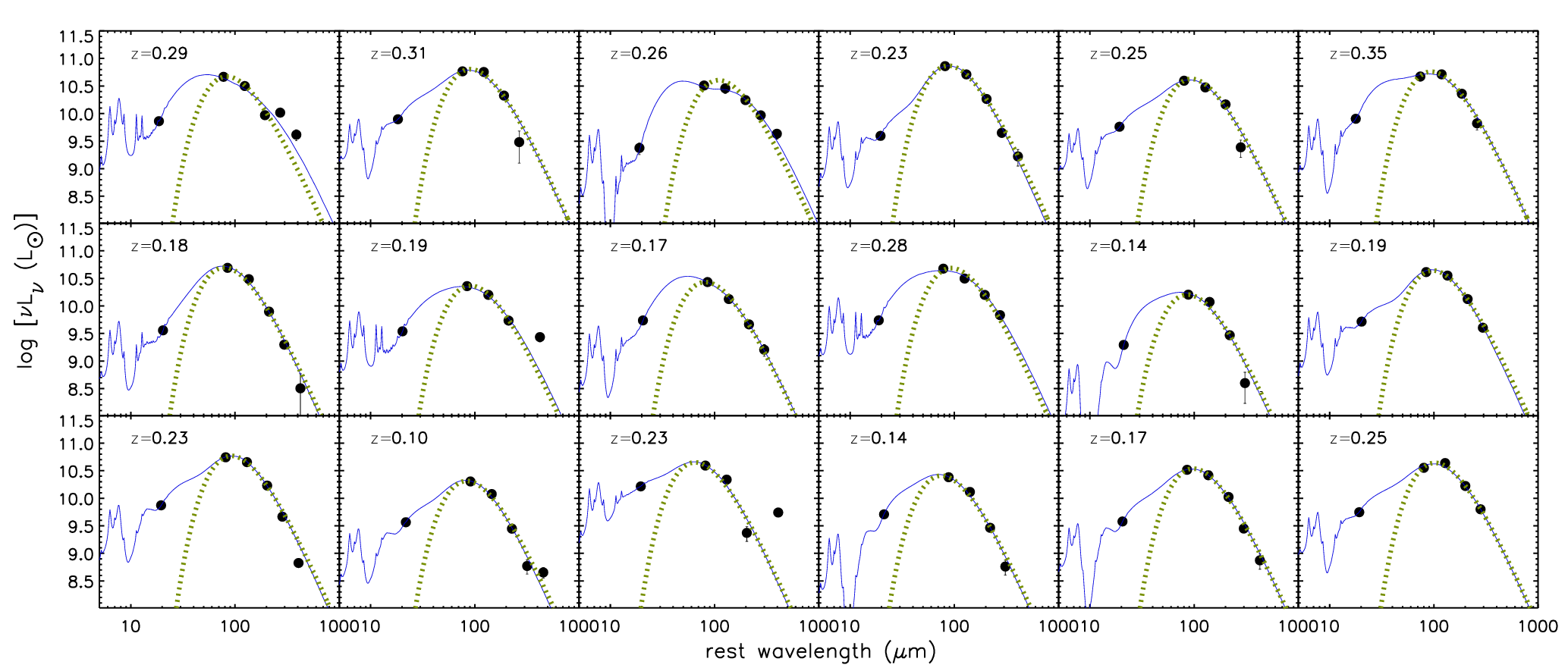,width=0.9\linewidth,clip=}\\
\epsfig{file=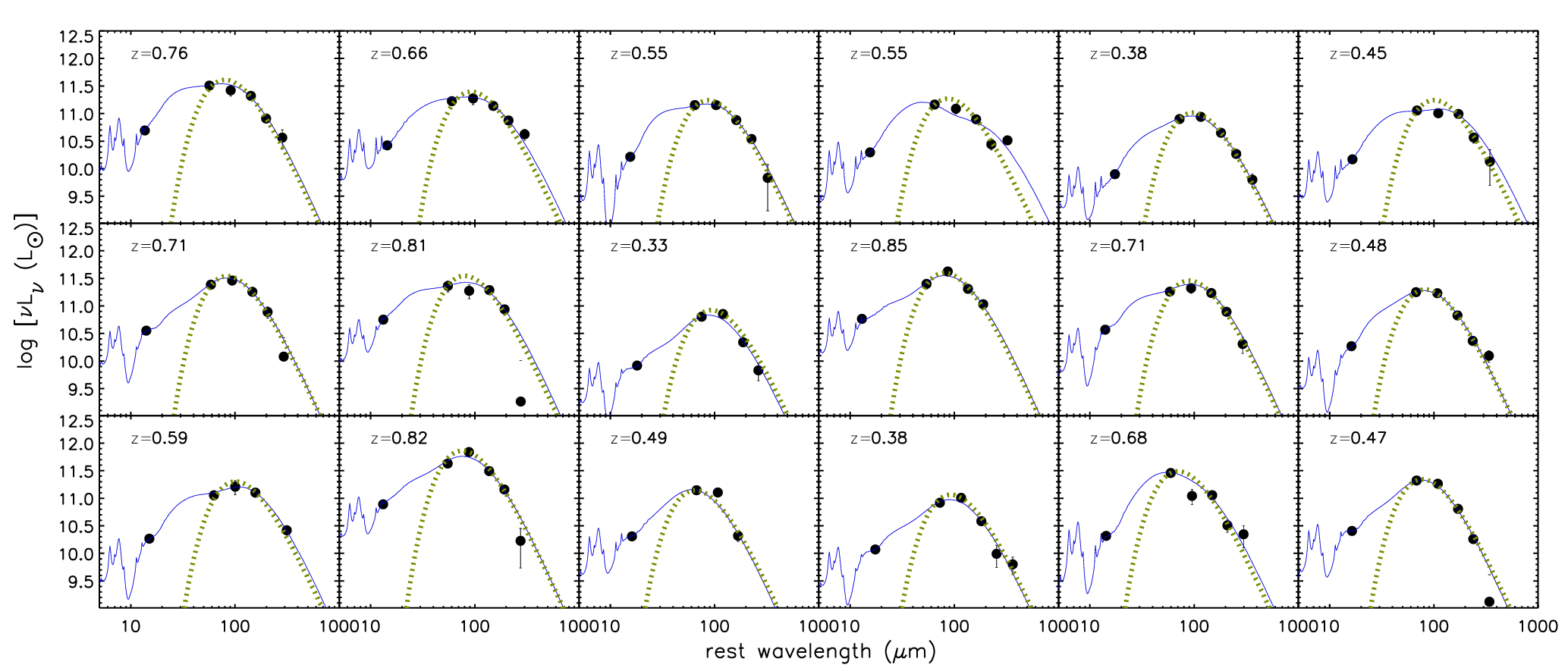,width=0.9\linewidth,clip=}\\
\epsfig{file=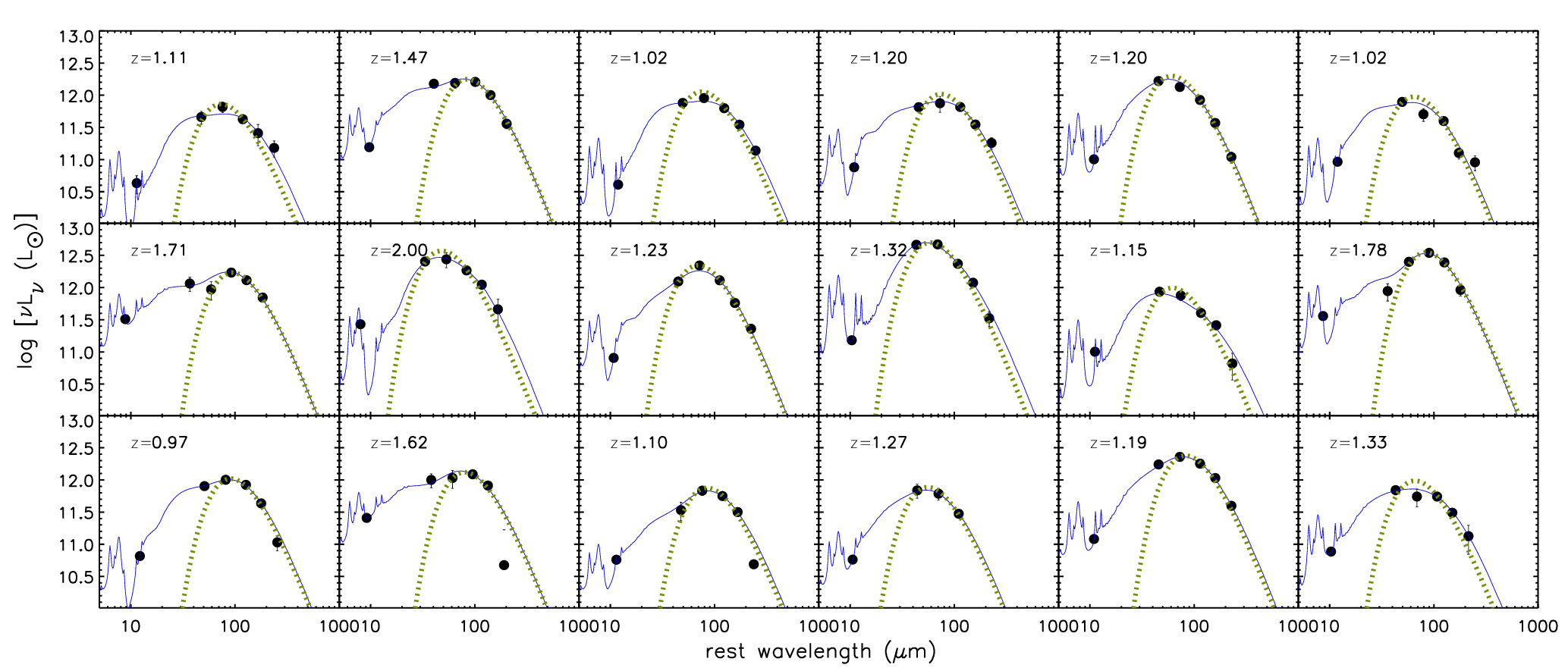,width=0.9\linewidth,clip=}\\
\end{tabular}
\caption{Typical SEDs for NIRGs, LIRGs and ULIRGs;
  y-axis rest-frame luminosity in L$_{\odot}$, x-axis rest-frame
  wavelength ($\mu$m). Black
  points are the available photometry from 24 to 500\,$\mu$m. The blue
  curve is the best-fit SK07 model and the green dotted curve is a
  single temperature greybody fit around the photometric peak in $\nu\,f_{\nu}$. } 
\label{fig:SEDs}
\end{figure*}

\subsection{SED characteristics}
\label{sec:seds}

Typical SEDs for the \textit{Herschel} sample ($0.1<z<2$) are shown in Fig.
\ref{fig:SEDs}, split into the 3 standard luminosity classes: normal IR
galaxies (NIRGs; 10$^{10}$$<$$L_{\rm IR}$$<$10$^{11}$), luminous IR
galaxies (LIRGs; 10$^{11}$$<$$L_{\rm IR}$$<$10$^{12}$) and
ultraluminous IR galaxies (ULIRGs; 10$^{12}$$<$$L_{\rm
  IR}$$<$10$^{13}$); see section \ref{sec:measurements} for details on
the SED fitting. The \textit{Herschel} bands cover the
bulk of the far-IR emission for the majority of sources, although in
some cases the exact position of the SED peak might be underestimated
or the slope of the mid-IR continuum might not be well constrained due
to lack of data between 24\,$\mu$m and the first \textit{Herschel}
band used in the fitting ($\lambda_{\rm obs}$=100 or 160\,$\mu$m). Nevertheless, in all cases the greybody function covers the bulk of
the dust emission giving a good representation of the average temperature
of the sources.

\begin{figure}
\centering
\epsfig{file=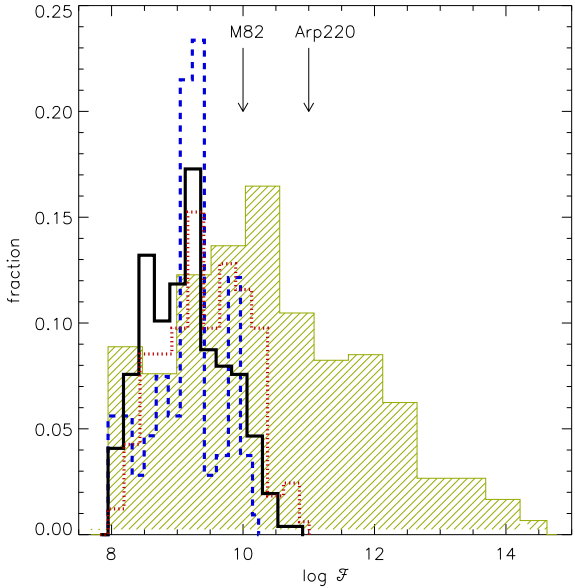, width=0.99\linewidth}
\caption{The distribution in $\mathcal{F}$ (log\,[$L_{\rm tot}$/4$\pi R^2$] in units of L$_{\odot}$\,kpc$^{-2}$) for NIRGS (blue dashed line), LIRGs (black line) and
  ULIRGs (red line). The hatched green region indicates the
  range covered by the SK07 library. The arrows indicate the value of
  $\mathcal{F}$ for M82 and Arp220 (taken from SK07).}
\label{fig:SK07_params2}
\end{figure}

To quantitatively describe the global SED shape we use the
$\mathcal{F}$ parameter defined in section \ref{sec:measurements}. Fig. \ref{fig:SK07_params2} shows the distribution in $\mathcal{F}$ 
for the \textit{Herschel} sample, split into the 3 luminosity classes and overlaid on the distribution of all
templates in the SK07 library. 
Interestingly, the $\mathcal{F}$ distributions of 
NIRGs, LIRGs and ULIRGs show large
overlap, perhaps surprising as one might expect ULIRGs to
exhibit a noticeable offset to larger $\mathcal{F}$ values, simply because they are more
luminous. However this is not the case, indicating that many of the
\textit{Herschel} ULIRGs are described by cool/extended rather than
warm/compact SEDs, in order to
reach the same radiation strength per unit area as their lower
luminosity counterparts. Indeed, the majority
of objects have 8.5$<\mathcal{F}<$10, suggesting that the IR-luminous population up to $z \sim$2 is
best described by extended rather than compact dust emission.
Note that the $\mathcal{F}$ distribution of the \textit{Herschel} sample covers only a small
range of the available parameter space. We find that templates with $\mathcal{F}>$11 
are not representative of any object (within the 1$\sigma$
uncertainties on $\mathcal{F}$) and in fact, only about a 1/3 of the number of templates in the
SK07 library are representative of the sample. This provides useful insight on what SED types are observationally confirmed in the context of a physically-motivated suite of
models.

\begin{figure}
\centering
\epsfig{file=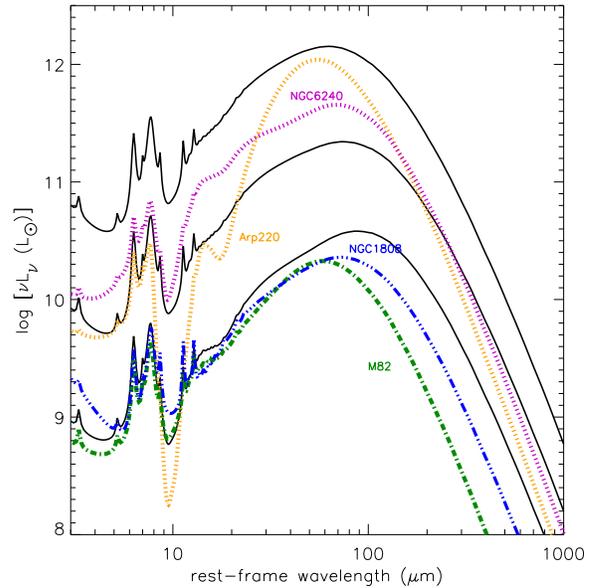, width=0.99\linewidth}
\caption{Average SEDs for the \textit{Herschel} sample (black curves) for NIRGs, LIRGs
 and ULIRGs, from bottom to top respectively. For comparison we also show the SEDs
 of Arp220, NGC62040, M82 and NGC1808 taken from SK07.}
\label{fig:average_seds_comp}
\end{figure}

Fig. \ref{fig:average_seds_comp} illustrates that the observed
distribution in $\mathcal{F}$ translates to broad-peaked SEDs
for the \textit{Herschel} sample, particularly evident when
comparing to those of well-studied local galaxies such as NGC1808, M82, NGC6240 and
Arp220 (their SEDs all taken from SK07). For the \textit{Herschel} SEDs, the slope on either side of the peak is shallow
in antithesis with the SEDs of the more compact starbursts M82 and
Arp220, which have $\mathcal{F}$=10 and 11 respectively.

\subsection{Far-IR colours}
\label{sec:colours}

\begin{figure*}
\centering
\epsfig{file=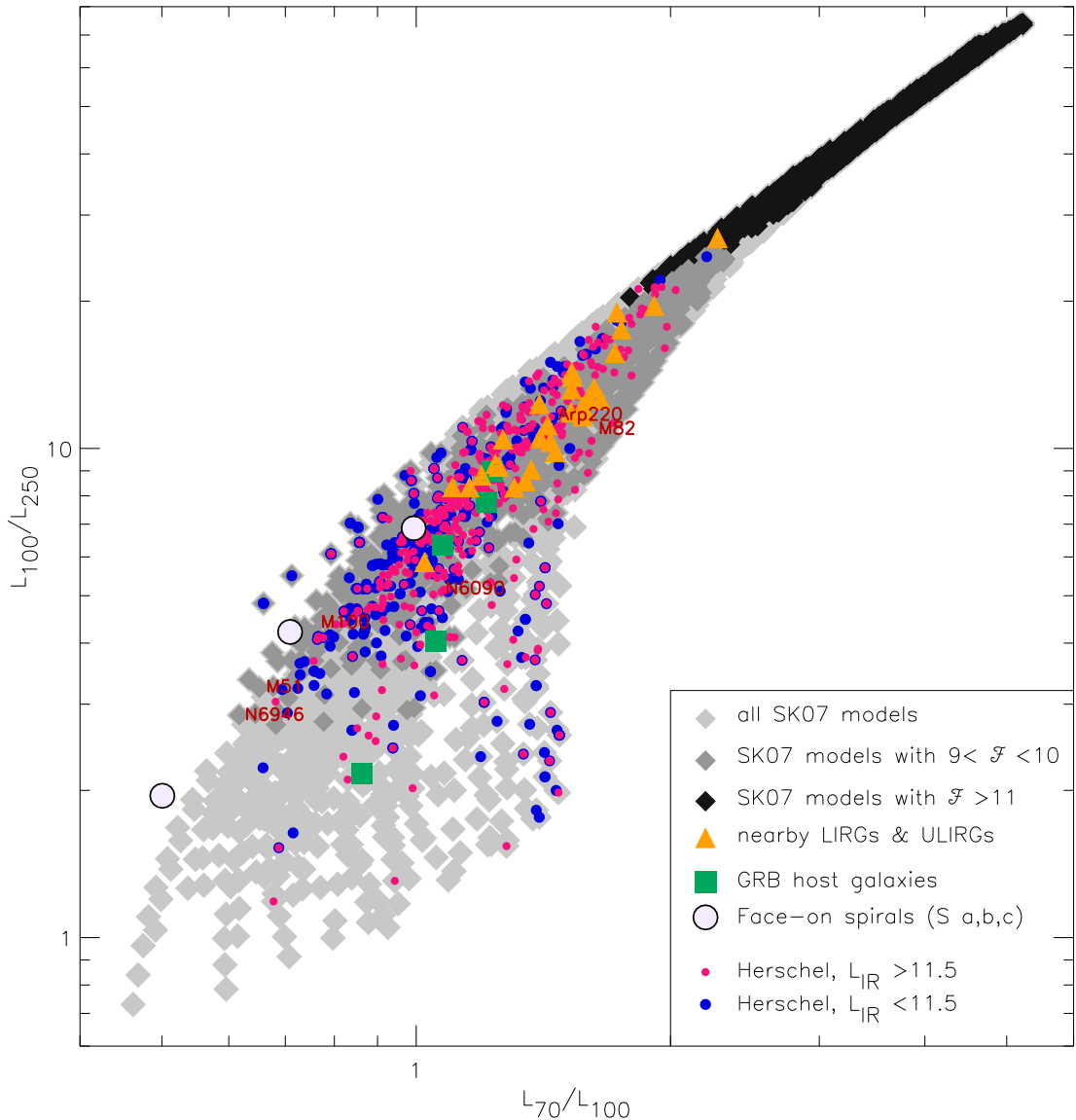, width=0.9\linewidth}
\caption{Rest-frame $L_{100}$/$L_{250}$ versus $L_{70}$/$L_{100}$
  colours for the \textit{Herschel} sample compared to other star-forming
  galaxy types indicated in the legend. The light grey shaded region represents
  the colours for the SK07 library, with dark grey shading used
  for templates of $9<\mathcal{F}<10$ and black shading for templates
  with $\mathcal{F}>11$. The location of face-on spirals (large white
  circles with black border) shifts from top-right to bottom-left with consecutive
morphological class, with Sa being in the top-right. The red writing
indicates the position of the well-known galaxies M100, M82, M51, NGC\,6946, Arp220 and NGC\,6090. }
\label{fig:colour}
\end{figure*}

Fig. \ref{fig:colour} shows the rest-frame $L_{100}$/$L_{250}$
versus $L_{70}$/$L_{100}$ colours of the best fit SK07 models to the \textit{Herschel}
sample. These bands were chosen for two reasons: (i)
they probe a part of the SED that is well sampled by our data hence substantially
constraining the SK07 models in that region and (ii)
they trace the shape of the SED both around the peak ($L_{70}$/$L_{100}$) and in the Rayleigh-Jeans side of the continuum
($L_{100}$/$L_{250}$). For comparison we also include the colours of
the SK07 library as well as those of nearby
galaxies and modelled SEDs computed using the GRASIL code (Silva
et al. 1998; 1999)\footnote{http://adlibitum.oat.ts.astro.it/silva/grasil/modlib/modlib.html}. These are: M100, M82, M51, NGC 6946,
Arp220 and NGC 6090, nearby LIRGs and ULIRGs (Vega et
al. 2008\nocite{Vega08}), modelled colours to represent face-on spirals (Sa, Sb and Sc), as well
as high redshift gamma-ray burst (GRB) host galaxies (Micha{\l}owski
et al. 2008\nocite{MIchalowski08}), in
essence young, compact, star-forming systems of low metallicity.
Nearby LIRGs and ULIRGs, such as M82 and Arp220,
show warm colours, whereas M51, M100 and M6946 are in the cold part of
colour-colour space. On the other hand, GRB hosts have similar values of
$L_{100}$/$L_{250}$ but warmer $L_{70}$/$L_{100}$ colours
than spiral galaxies. Note that the modelled spirals (Sa, Sb, Sc) are located along
the edge of the available parameter space with colours that become colder down the
sequence from Sa to Sc. Although the colours of the Sc spiral are
slightly offset from the parameter space that the SK07 templates
cover, small discrepancies between SED libraries are expected,
particularly since there are many input parameters which can contribute to
the final SED shape.

\begin{figure*}
\centering
\epsfig{file=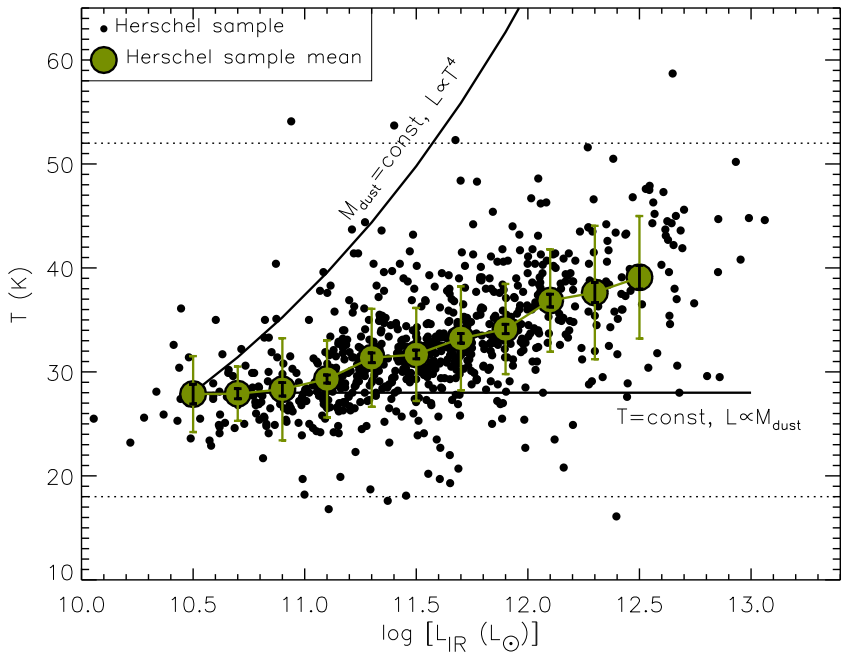, width=0.7\linewidth}
\caption{Dust emperature as a function of total infrared
  luminosity for the \textit{Herschel}
  sample (black points). The green filled circles are the mean temperatures computed for 11 $L_{\rm IR}$ bins (see table
  \ref{table:temperature}). The green error bars are the standard
  deviation in the measured temperatures of each bin, whereas the
  black error bars are the error on the mean. 
The 2 limiting scenarios
  of the Stefan-Boltzmann law are also shown (solid black lines). The
  dotted horizontal lines outline the $L-T$ parameter space in which
  we are complete; see section \ref{sec:herschel_selection}.}
\label{fig:temperature_herschel}
\end{figure*}

Overall, the SK07 templates extend over a large range in colour-colour
space, adequately covering the colours of the comparison sample. 
The SK07 colours show a large spread below $L_{100}$/$L_{250} \sim$10 and a narrow
tail at large values of $L_{100}$/$L_{250}$ and
$L_{70}$/$L_{100}$. This tail, formed by models with $\mathcal{F}>11$, is not populated by any of the galaxy
groups presented here, suggesting that such hot SEDs are not typical
of any type of star-forming galaxy. 
The darker grey shaded region consists of
templates with $9<\mathcal{F}<10$ and is the parameter
space that most \textit{Herschel} sources occupy
(see Fig. \ref{fig:SK07_params2}). These
templates have cool far-IR $L_{100}$/$L_{250}$ colours of
$<$10 and a large spread in $L_{70}$/$L_{100}$. A value of
$L_{70}$/$L_{100}$=1.0\,$\pm$0.3 ties in with an SED
peak between 70 and 100\,$\mu$m, the range seen in the \textit{Herschel} sample (see section
\ref{sec:temperature}). 
The light grey lower-left part of colour-colour space is scarcely occupied by the
\textit{Herschel} sample, indicating that very cold, cirrus-dominated SEDs
with $L_{100}$/$L_{250} <$3  are uncommon in $L_{\rm IR}>10^{10}$\,L$_{\odot}$ galaxies. This is not a
consequence of either the selection or the survey flux limits; we
remind the reader that our sample is complete with respect to the SED
types that can be probed, hence a deeper IR survey is not expected to identify
IR-luminous galaxies with different SEDs to the ones observed
here. Colder colours ($L_{100}$/$L_{250}<$3) might perhaps be more
common amongst more quiescently star-forming galaxies with lower
infrared luminosities ($L_{\rm IR}<10^{10}$\,L$_{\odot}$), but
examining such sources is beyond the scope of this paper.
 
It is interesting to compare the locus of
\textit{Herschel} sources with that of the comparison samples. We see
significant overlap overall, however many
\textit{Herschel} LIRGs and ULIRGs are clearly offset from the
region traditionally occupied by their local counterparts, displaying 
colder colours consistent with more quiescent star-forming galaxies
such as M100 and NGC\,6090.

\begin{table}
\centering
\caption{The mean temperature, 1$\sigma$ scatter per $L_{\rm IR }$
  bin and error on the mean for the \textit{Herschel} sample --- see Fig. \ref{fig:temperature_herschel}. The last column is the median redshift of each $L_{\rm IR}$ bin.}
\begin{tabular}{l|c|c|c|c|}
\hline 
log\,$L_{\rm IR}$ & mean $T$ & 1$\sigma$ & error
on & median $z$\\
 (L$_{\odot}$) & (K) &(K)  &mean $T$ (K) & \\
\hline
   10.4--10.6   &    27.9   &     3.7 & 0.94& 0.14\\
    10.6--10.8   &  27.9  &      2.6 & 0.48 &0.19\\
   10.8--11    & 28.3    &    4.9 &   0.64&0.23\\
   11--11.2     &  29.3    &    3.7 & 0.39 &0.36\\
   11.2--11.4&       31.4   &     4.7 &0.47  &0.42\\
   11.4--11.6&       31.7   &     4.5 & 0.42 &0.54\\
   11.6--11.8&       33.2   &     5 &   0.46&0.67\\
   11.8--12&       34.1  &      4.3 &  0.45&0.84\\
  12--12.2&       36.9   &     4.9 & 0.62 &0.94\\
  12.2--12.4&       37.6  &      6.4 &  0.97 &1.09\\
  12.4--12.6&       39.1  &      5.9 &1.13  &1.23\\
\hline
\end{tabular}
\label{table:temperature}
\end{table}

\begin{figure*}
\centering
\epsfig{file=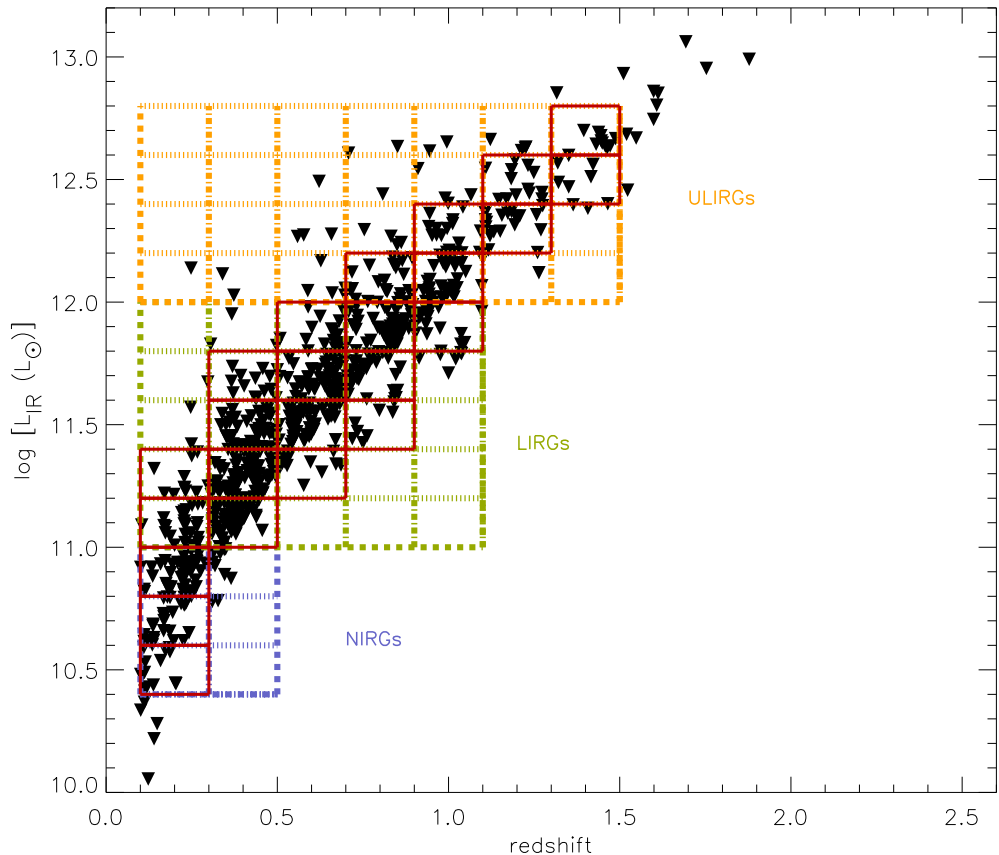, width=0.8\linewidth}
\caption{The luminosity-redshift parameter space of the final
  \textit{Herschel} sample
  used in the analysis (black triangles), split into bins of
size 0.2 dex in luminosity and 0.2 in redshift. The bins are outlined
in blue for NIRGs, green for LIRGs and orange for ULIRGs and further
delineated in
red if used when computing the average SED shape in that bin (see
Fig. \ref{fig:average_seds} for average SEDs). }
\label{fig:LIRz_finalsample}
\end{figure*}

\subsection{The luminosity - dust temperature relation}
\label{sec:temperature}

The dust temperature ($T$) for the \textit{Herschel} sample, derived
as described in section \ref{sec:measurements},  is shown in
Fig. \ref{fig:temperature_herschel}  as a function of total infrared
luminosity. Note that although the derivation of single average dust
temperatures represents simplistic assumptions with respect to dust properties, such as
optical depth, emissivity, dust geometry and so on, it is currently
the only consistent way to characterise and compare statistically
large samples of IR-luminous galaxies over a large redshift range. 

The mean temperature of the \textit{Herschel} sample ranges from about 28 to
39\,K, increasing with $L_{\rm IR}$ and showing an average 1\,$\sigma$
scatter of 5\,K; see table \ref{table:temperature}. Our results confirm that the choice of
temperature range ($\sim$18--52\,K) over which our sample was
described as unbiased (section \ref{sec:selection}) was adequate, as we find that the minimum and maximum average
temperatures are offset by about 10\,K from 18 and 52\,K
respectively. In fact, IR-luminous galaxies with $T<$25\,K
and $T>$45\,K constitute $\sim$6 and $\sim$3 per cent of the total
population respectively. 

Since the emission from large dust grains in equilibrium, hence the bulk of the
IR emission, is well approximated by the black body (or grey body) function, we also investigate to what extent we can
use the Stefan-Boltzmann law, $L=A\epsilon\sigma T^4$, to interpret
the $L-T$ relation, where $A$ is the surface area, $\epsilon$ is the
emissivity and $\sigma$ is the Stefan constant. A is proportional to
$R^2$, which is in turn proportional to the dust mass ($M_{\rm dust}$),
for constant extinction, so one can re-write the
Stefan-Boltzmann law as $L\propto M_{\rm dust} T^4$ (or $L\propto R^2 T^4$). This spawns two limiting scenarios. The
first is that the emitting area and/or dust mass
is constant which would result in an $L-T$
relation of the form: $L\propto T^4$. The second is
that the emitting area and/or dust mass is proportional to the luminosity
($L\propto R^2$ or $L\propto M_{\rm dust}$) with the temperature remaining
constant for all galaxies. The curves representing these scenarios are
plotted in Fig. \ref{fig:temperature_herschel}.
The observed $L-T$ relation for the \textit{Herschel} sample is
quite flat--- 2 orders of magnitude increase in luminosity results in only a 40 per cent
increase in temperature --- and hence closer to the $L\propto M_{\rm
  dust}$ limiting scenario. This suggests that the $L-T$ relation is mainly
shaped by an increase in dust mass and/or IR emitting radius and less
so by an increase in dust heating. In
other words, the average dust temperature of ULIRGs is much lower than
what one would expect if their increased luminosity were the only
factor shaping the $L-T$ relation. This indicates that the
dust masses and/or sizes of ULIRGs are larger than those
of NIRGs, significantly diluting the effect that their increased luminosity has on the
temperature.

\begin{figure*}
\centering
\epsfig{file=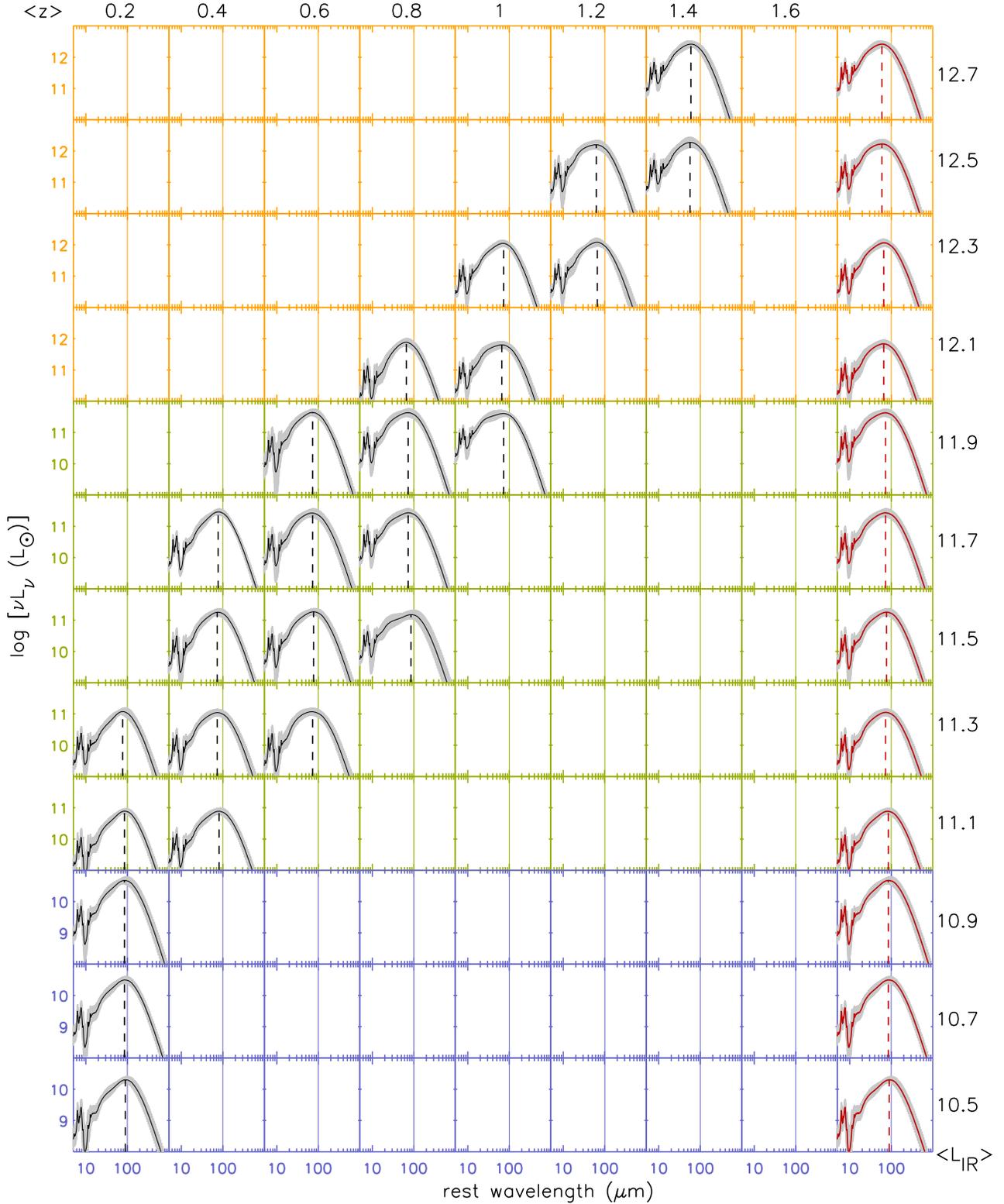, width=0.99\linewidth}
\caption{Average SEDs ($\nu L _{\nu}$ vs $\lambda_{\rm rest}$) for
  the \textit{Herschel} sample. This figure is
analogous to Fig. \ref{fig:LIRz_finalsample}, such that each row represents a
change in redshift interval, and each column a change in luminosity
interval. The central luminosity and redshift of each bin are shown on
the left side and top of the plot respectively. Average SEDs are shown for $L-z$ bins with 5
  or more objects, outlined in red in Fig. \ref{fig:LIRz_finalsample}. The boxes are coloured blue for NIRGs, green for
  LIRGs and orange for ULIRGs. The solid vertical line in the middle of each
  box is at 100\,$\mu$m, whereas the dashed line denotes the SED
  peak. The red SEDs at the end of each row are the average SEDs for
  that row. }
\label{fig:average_seds}
\end{figure*}

\begin{figure*}
\centering
\begin{tabular}{c|c|}
 \epsfig{file=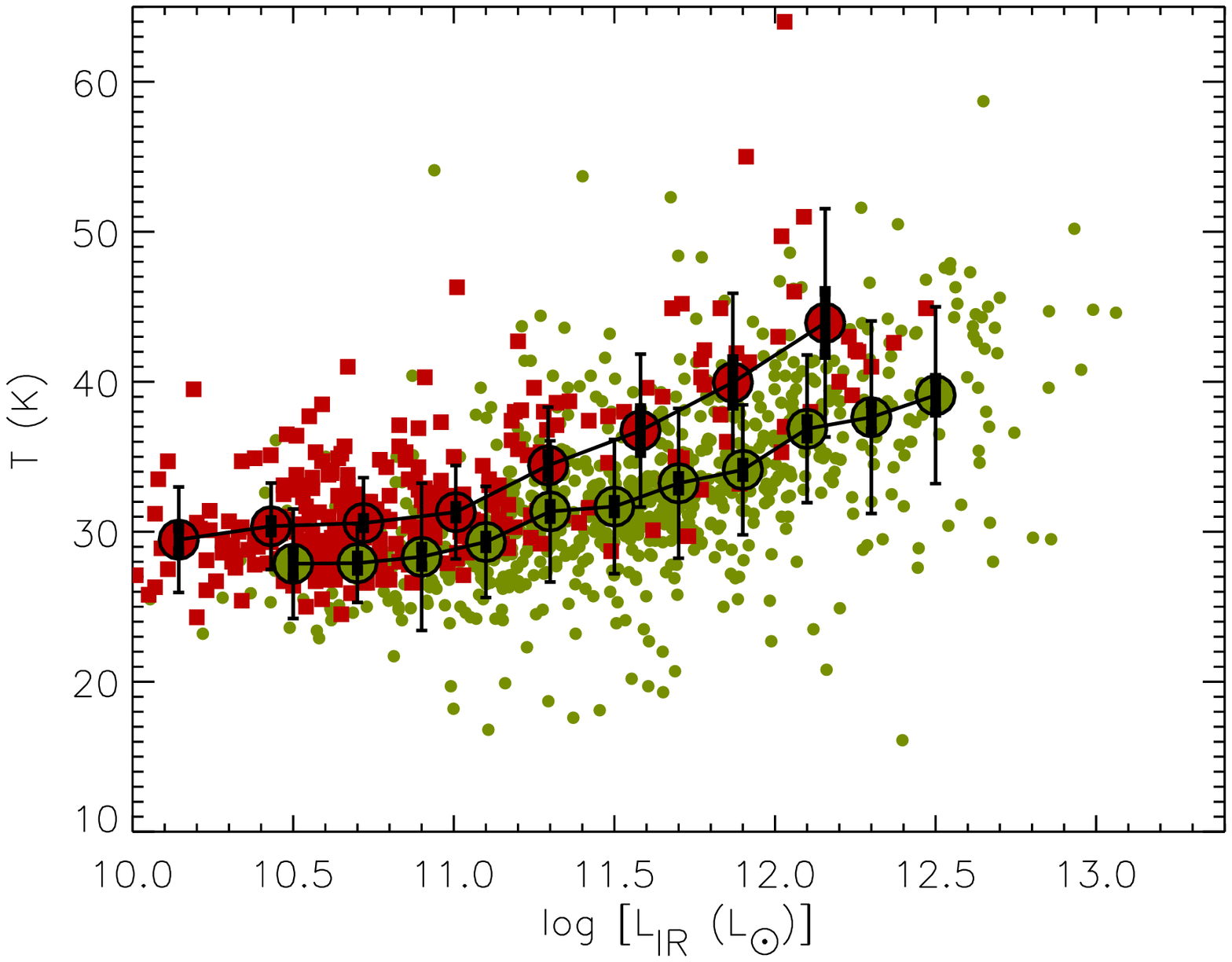,width=0.49\linewidth,clip=} & \epsfig{file=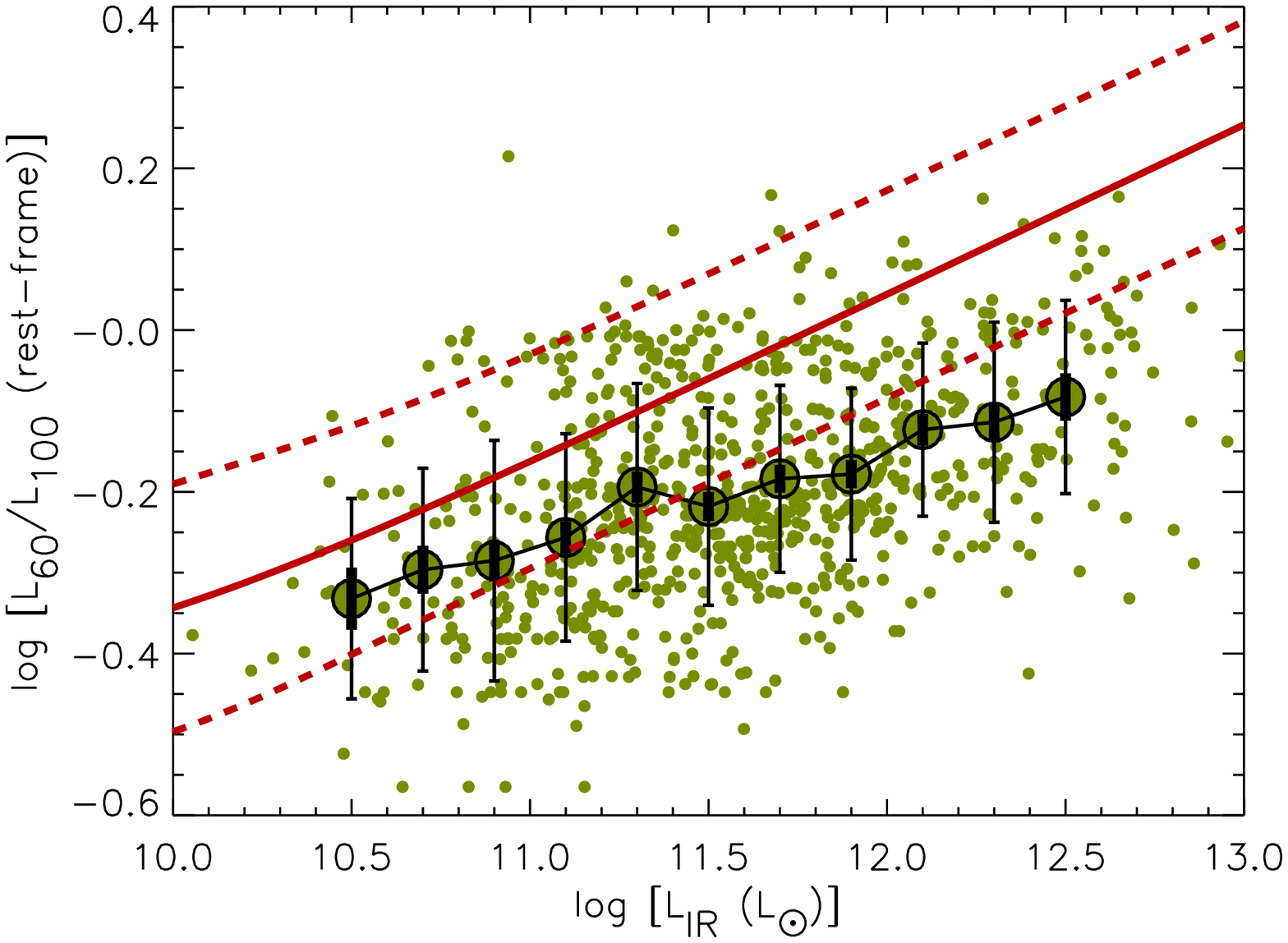,width=0.49\linewidth,clip=} \\
\end{tabular}
\caption{\textit{Left panel}: Dust temperature vs $L_{\rm IR}$ for the
  \textit{Herschel} sample (green circles) and the local ($z\lesssim$0.1) sample (red
  squares). The mean trends for the two samples are also shown (large
  filled circles) with the large thin error bars representing the
  1$\sigma$ scatter in each bin and the short thick error bars
  representing the error on the mean. \textit{Right
  panel}: rest-frame colour (L$_{60}$/L$_{100}$) as a function of total infrared
luminosity for the \textit{Herschel}
sample (green circles). The large filled circles represent the mean
trend with the large thin error bars representing the
  1$\sigma$ scatter in each bin and the short thick error bars
  representing the error on the mean. The solid and dashed lines are the local luminosity-colour
relation and 1$\sigma$ limits from Chapin, Hughes $\&$ Aretxaga
(2009).} 
\label{fig:temperature}
\end{figure*}

Fig. \ref{fig:LIRz_finalsample} shows the $L-z$ distribution of the
sample, split into bins of 0.2\,dex in
luminosity and 0.2 in redshift. For the bins additionally outlined in
red (24 in total), containing $\ge$\,5 objects, we compute the average
SED in that bin shown in
Fig. \ref{fig:average_seds}. Fig. \ref{fig:average_seds} is
analogous to Fig. \ref{fig:LIRz_finalsample}, such that each row represents a
change in redshift interval, and each column a change in luminosity interval. The shaded
region represents the
1$\sigma$ scatter around the average SEDs, whereas the red SED at the end of each row is the
average for that row. Consistent with what we observe with regard to
the $L-T$ relation (Fig. \ref{fig:temperature_herschel}),
Fig. \ref{fig:average_seds} demonstrates that there is a shift in
the SED peak from longer to
shorter wavelengths with increasing infrared luminosity. This is more clear in the last column which shows the average
SED for each luminosity bin: the SED peak ($\lambda_{\rm peak}$) shifts from
86\,$\mu$m in the lowest luminosity bin to 65\,$\mu$m in the highest
luminosity bin. This is also seen in
Fig. \ref{fig:average_seds_comp} where the average SED of each
luminosity class is shown, with the mean and standard deviation in $\lambda_{\rm peak}$ being 86$\pm$18\,$\mu$m for NIRGs,
75$\pm$18\,$\mu$m for LIRGs and 65$\pm$17\,$\mu$m for ULIRGs. Note
that the 1\,$\sigma$ scatter is large, partly because the peak is not always
well constrained by our data, and partly because of
the large diversity in SED types (see Figs \ref{fig:SEDs}
and \ref{fig:colour} for examples). 

Another feature that appears to change with $L_{\rm IR}$, is
the silicate absorption depth, becoming shallower for higher $L_{\rm
  IR}$. Our data does not probe the depth of the silicate feature,
except over a small window in redshift
where it coincides with the 24\,$\mu$m passband. Hence this trend is likely an artifact brought about by model
degeneracies. In the SK07 formulation, visual
extinction is tied in to the silicate absorption depth, the slope
of the mid-IR continuum and the SED peak wavelength. These
quantities are significantly degenerate at low redshifts, although for
high redshift sources ($z \gtrsim $1), the photometry probes further into the mid-IR continuum, placing
additional constraints on the extinction. Nevertheless, although the
SED shape in the near/mid-IR is more reliably reproduced for
the high redshift sources, the observed trend of
decreasing silicate depth with increasing $L_{\rm IR}$ is most likely
artificial.

\subsection{Evolution in dust conditions}
\label{sec:evolution}

\begin{figure*}
\centering
\epsfig{file=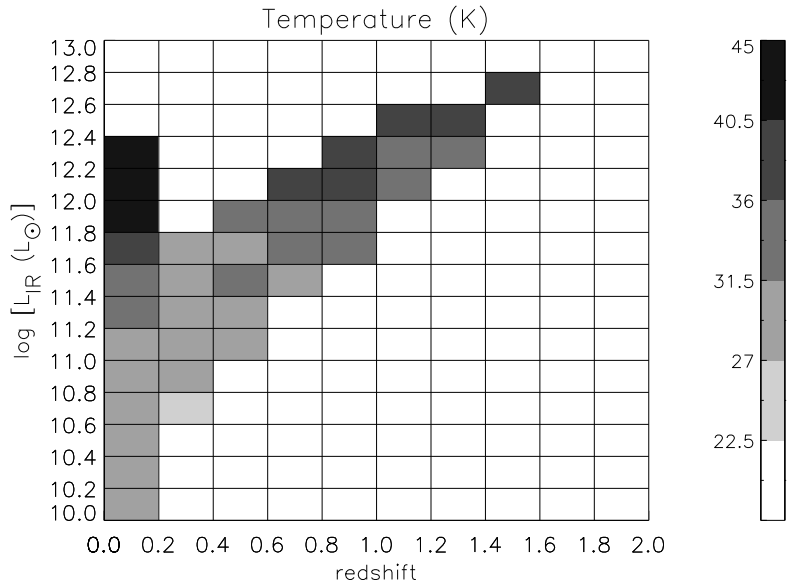 ,width=0.6\linewidth}
\caption{2-D image of the $L-T-z$ space for IR-luminous galaxies from
  the local Universe to $z=2$. This includes both the local and
  the \textit{Herschel} samples. Objects are divided
  into $L_{\rm IR}$ (y-axis) and $z$ bins (x-axis), and the average
  temperature of each bin is shown as a greyscale intensity map. The
  colourbar on the right is the temperature key for the map. The
  temperature bins extend from 22.5 to 45\,K. 
  White colour indicates unpopulated or underpopulated (i.e. $<$5
  objects) $L-T-z$ bins.}
\label{fig:image}
\end{figure*}

\begin{figure*}
\centering
\epsfig{file=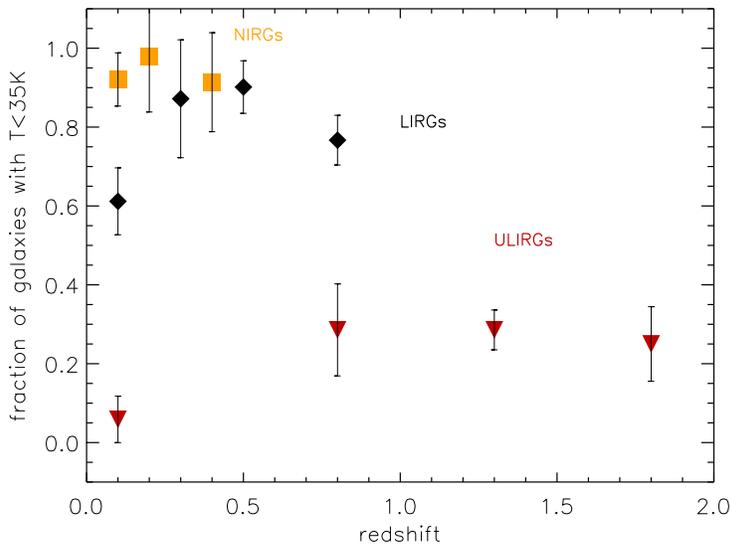, width=0.6\linewidth}
\caption{The fraction of $T$\,$<$35\,K NIRGs (orange squares), LIRGs (black diamonds) and ULIRGs
  (red triangles) as a function of redshift. The local sample is at
  $z\sim$0.1 and the remaining bins include sources from the
  \textit{Herschel} sample only.}
\label{fig:temp_z}
\end{figure*}

In Fig. \ref{fig:temperature}, we compare the properties of the
\textit{Herschel} sources to those of the local sample, assembled as described in section
\ref{sec:localsample}. The local luminosity-temperature and
luminosity-colour ($C$; $L_{\rm 60}/L_{\rm 100}$) relations, the latter in
functional form from Chapin, Hughes
$\&$ Aretxaga (2009\nocite{CHA09}), are shown in Fig. \ref{fig:temperature}, left and right
panels respectively. The $L_{\rm 60}/L_{\rm 100}$ colour has been used extensively to characterise
the dust temperature of local samples and analysis of \textit{IRAS}-selected galaxies
has shown that more luminous sources have higher colour temperatures
than their less luminous counterparts (e.g. Dunne et al. 2000\nocite{Dunne00};
Dale et al. 2001\nocite{Dale01}; Dale $\&$ Helou 2002\nocite{DH02};
Chapman et al. 2003\nocite{Chapman03}). 
For both $L-T$ and $L-C$ relations we note a systematic difference between the
\textit{Herschel} and local samples, with the former displaying lower
values of $T$ and $L_{\rm 60}/L_{\rm 100}$. This can be interpreted as evidence for evolution: high redshift IR-luminous galaxies have
more emission longward of $\sim$60\,$\mu$m compared to their low redshift analogues, lowering
the average dust temperature. This is consistent with results from
section \ref{sec:colours}, Fig. \ref{fig:colour}, where we see that
the \textit{Herschel} sample extends to colder far-IR colours than local
LIRGs and ULIRGs.

Recent results from the Planck collaboration (Part
16\nocite{Clements11}) on the dust properties of nearby
\textit{IRAS}-selected sources showed that many local $10^{10} <L_{\rm IR}<10^{11}$ galaxies
extended to lower temperatures than previously reported. This does not
affect our work as (i) within the scatter and given that emissivity is a free parameter in their
fitting, their measured dust temperature in the $10^{10} <L_{\rm IR}<10^{11}$ luminosity range are consistent
with the ones we report here for the local sample
(Fig. \ref{fig:temperature}) and (ii) as described in section
\ref{sec:localsample}, we assemble the local sample in an unbiased
part of parameter space. Furthermore, lack of additional submm/mm data
for our nearby sources, would not change the average dust temperatures measured,
as these trace the peak dust emission and are thus insensitive
to inclusion of photometry significantly longward of the peak (see also Magnelli et al. 2012).

Although a systematic reduction in the dust temperature
and colour of IR-luminous galaxies from the local to the high redshift
Universe is observed,
it is worth examining how this translates to changes in the SED shape
with redshift within the \textit{Herschel} sample. In Fig.
\ref{fig:average_seds}, we see some evidence for a shift in the SED peak to longer wavelengths
along the rows, i.e. with increasing redshift, for example in the log\,$L_{\rm IR}$=11.4--11.6
bin. However, overall, it is not clear how the SED shape evolves with redshift. This is likely
a consequence of sparse sampling of the SK07 templates by our photometry,
exacerbated by the small redshift range covered along each row,
statistical uncertainties in each bin as well as the fact that in many
cases, sources do not cover each $L-z$ bin uniformly (see
Fig. \ref{fig:LIRz_finalsample}). A different way to investigate
evolution in dust properties is to repeat this exercise with our computed
dust temperatures, in order to remove model-dependent uncertainties,
although the other sources of uncertainty outlined above would remain. 
Fig. \ref{fig:image} shows a 2-D image of the dust
temperature of IR luminous galaxies (local and \textit{Herschel} samples combined) as a function
of redshift and luminosity, in bins of 0.2 and 0.2\,dex
respectively. Besides an increase in average dust temperature vertically along the
luminosity axis, consistent with our analysis in section \ref{sec:temperature} on the
$L-T$ relation, we also note an overall reduction in the mean
dust temperature, horizontally, along the redshift axis. This is more
pronounced when considering the first and last bins of each row,
however in some cases it is also evident along the length of the
row. 

Considering the above trends, it is interesting to examine whether the
evolution in dust temperature we observe, is luminosity dependent,
i.e. whether dust conditions of ULIRGs evolve at a different rate
to those of NIRGs (Fig. \ref{fig:temp_z}). 
We place the separation between cold and warm at $T$=35K, corresponding to
the mean temperature for the \textit{Herschel} sample shown
in Fig. \ref{fig:temperature}. Fig. \ref{fig:temp_z}
shows an increase in the fraction of cold ULIRGs with redshift, from 5
per cent in the local Universe to about 30 per cent at $z\sim$1--2. Also the fraction of cold
LIRGs increases from about 60 per cent locally to about 80 per cent
at high redshift. However, as for Figs \ref{fig:average_seds} and
\ref{fig:image}, in some bins, the $L-z$ parameter space is not sampled
uniformly. The higher redshift bins in Fig. \ref{fig:temp_z} include a larger fraction of more luminous
and hence warmer sources, implying that the fraction of cold sources is likely
underestimated, resulting in the observed downturn of the computed
fraction. 

Note that given the depth of our data, currently this is the best attainable coverage of the $L-z$
plane in the horizontal ($z$) direction within the unbiased
framework we define in this work. This does not mean that LIRGs at $z\gtrsim1$ and ULIRGs at $z\gtrsim2$
will not be detected by \textit{Herschel} or other
facilities, rather it implies that it is not currently possible to measure the \textit{aggregate} properties of the
LIRG and ULIRG population at those redshifts. On the other hand, with larger area
\textit{Herschel} surveys we expect to cover the gap between the local and high redshift
Universe in the vertical ($L$) direction, enabling better
sampling of the $L-z$ plane, where large survey area is needed, and hence
achieve better statistics for log\,$L_{\rm IR}$/L$_{\odot}$\,$>$11.5
galaxies up to $z=1$. 

With respect to the selection at 24\,$\mu$m, we saw that
we are likely missing a few per cent of SEDs with steep far-to-mid-IR
continua and deep silicate features, about half of which have
$f_{500}$/$f_{160}$\,$>$1 extending to higher values than we find in the
\textit{Herschel} sample (see Fig. \ref{fig:ratio160_24} and section \ref{sec:24um_selection}). This
suggests that many of the sources missed by the 24\,$\mu$m
criterion might have lower temperatures than the average temperatures
derived for the \textit{Herschel} sample. Although this fraction of
sources is very small and hence unlikely to change our results, it would
nevertheless only serve to strengthen the differences we observe
between the local and high redshift sample.

\section {Discussion}
\label{sec:discussion}

\subsection{The properties of the IR-luminous population}
The \textit{Herschel} sample under study consists mainly of LIRGs (64
per cent), with ULIRGs constituting 20 per cent, consistent with what
is expected in the redshift range probed ($0.1<z<2$), whereas the remaining 16 per
cent are normal IR galaxies (NIRGs;  $10^{10} <L_{\rm
  IR}<10^{11}$\,L$_{\odot}$). The dust temperatures of the sample show large scatter from
15 to 55\,K, however the mean temperature ranges from 28$\pm$4 to
39$\pm$6\,K, with ULIRGs being on average about 10\,K warmer than
NIRGs. Similarly, we see a shift in the average
SED peak wavelength from 86$\pm$18\,$\mu$m for NIRGs to 65$\pm$17\,$\mu$m for
ULIRGs. The sample is best described by cool/extended, rather than
warm/compact SEDs, and broad peaks, translating to low values of
$\mathcal{F}$, a parameter which we defined as a measure of the
overall IR SED shape. In addition, there is a large overlap in colour--colour
($L_{100}$/$L_{250}$ -- $L_{70}$/$L_{100}$) space
between the \textit{Herschel} sample and other star-forming galaxy types, the
coldest of which (spirals and young compact star-forming galaxies) are at
$L_{100}$/$L_{250}$\,$\lesssim$7 and the warmest (ULIRGs and
starbursts) at $L_{100}$/$L_{250}$\,$\gtrsim$7. 
For the \textit{Herschel} sample, we noted a roughly equal number of sources above and
below $L_{100}$/$L_{250}$\,$\sim$7. 
It is worth mentioning that only about a 1/3 of the SK07 templates
are representative of the \textit{Herschel} sample. The
$L_{100}$/$L_{250}$$>$20, $L_{100}$/$L_{250}$$<$3 and $L_{70}$/$L_{100}$$>$2 regions of the SK07 library are scarsely
populated. Moreover, the distribution in $\mathcal{F}$ of the
SK07 library extends to much higher values ($\mathcal{F}$\,$\sim$15), than
what is observed in the \textit{Herschel} sample ($\mathcal{F}$\,$<$11). This indicates that compact, hot-dust-dominated
SEDs or very cold cirrus-dominated SEDs are
not typical of the IR-luminous population. 

How do our findings compare to other studies of IR galaxies?
As mentioned earlier, we have performed a rigorous analysis of selection
effects in order to minimise any biases which would interfere with our results. Hence, due to the nature of our study, we are sensitive
to most, if not all, IR galaxy types with $L_{\rm
  IR}$\,$>$\,10$^{10}$\,L$_{\odot}$ at $z$=0.1--2. Consequently, results on the properties of IR galaxies from previous
studies with \textit{Spitzer} (e.g. Symeonidis et al. 2009; Kartaltepe
et al. 2010a; Patel et al. 2011) and SCUBA (e.g. Kovacs et al. 2006; Coppin et
al. 2008; Santini et al. 2010\nocite{Santini10}) are all within the parameter space we probe (see also
Magnelli et al. 2012). The results reported recently using
\textit{Herschel} data (e.g. Rowan-Robinson et al. 2010; Hwang et
al. 2010; Smith et al. 2012\nocite{Smith12}) are also within the parameter space we
define here for the IR-luminous population. However, we note that although
cold, cirrus-dominated SEDs such as those reported in
Rowan-Robinson et al. (2010) and Smith et al. (2012) are part
of our sample, they represent a small fraction ($<$6 per cent) of the
IR-luminous population and are not the prevalent SED
types.

\subsection{The $L-T$ relation}
The increase in dust temperature as a function of infrared luminosity
we observe here, is similar to the $L-T$ relation that has emerged
from most infrared population studies (e.g. Dunne et al. 2000\nocite{Dunne00}; Dale et al. 2001\nocite{Dale01}; Dale $\&$
Helou 2002\nocite{DH02}; Chapman et al. 2003\nocite{Chapman03}). 
Since the emission from large dust grains in equilibrium, hence the bulk of the
IR emission, is well approximated by a black (or grey) body function,
we aimed to understand the $L-T$ relation within the framework of the Stefan-Boltzmann law. As described in section
\ref{sec:temperature}, the two limiting scenarios
of the Stefan-Boltzmann law are: (i) $L\propto T^4$ with $R$ (the radius of the emitting
region), or $M_{\rm dust}$ (the dust mass) kept constant and
(ii) $L\propto R^2$ or $L\propto M_{\rm dust}$ with $T$ kept constant. The former scenario would produce a steep $L-T$
relation, whereas for the latter the $L-T$ relation would be flat,
with the increase in luminosity tying in with an increase in surface
area of the emitting body or the dust mass. 
We find that the $L-T$ relation for the \textit{Herschel} sample lies
closer to the latter scenario suggesting that its shape is mainly driven by an increase in dust
mass or extent of dust emitting region and less so by energetics. In
other words, it seems that the increased dust heating in ULIRGs is
diluted by an increase in their physical size and/or
dust mass, such that the $L-T$ relation is diverted away from the
$L\propto T^4$ scenario. 
However, although more strongly star-forming galaxies such as ULIRGs are predominantly more massive
(e.g. Dav\'{e} 2008\nocite{Dave08}; Shapley 2011\nocite{Shapley11}) and dustier
systems (e.g. Magdis et al. 2012\nocite{Magdis12}) with warmer average dust temperatures, we find that their SED shapes are
not substantially different to their lower luminosity counterparts. In
particular, we do not see a clear segregation in far-IR colour-colour
space or in the range covered by
$\mathcal{F}$ as a function of infrared luminosity. Interestingly, this suggests that properties such as optical depth,
dust extinction, extent of IR emitting regions etc., which determine the
overall SED shape are not substantially different between high
redshift ULIRGs and their lower luminosity systems.

\subsection{Evolution of the IR-luminous population}

In this work we found evidence that the dust temperatures of
\textit{Herschel} sources are systematically colder than equivalently
luminous galaxies in the local Universe. The rigorous analysis of survey biases we performed ensures the
validity of this result, as our final sample selection was sensitive to all IR galaxies with $18 \lesssim T (\rm K)
\lesssim 52$. We found sources at $z>$0.5
and log\,[$L_{\rm IR}$/L$_{\odot}$]\,$>$11.0 to be on average
between 5 and 10\,K colder than their $z<$0.1
counterparts. We also noted a systematic offset to colder rest frame
$L_{60}$/$L_{100}$, $L_{100}$/$L_{250}$ and $L_{70}$/$L_{100}$ colours
for the \textit{Herschel} sample in comparison to local equivalent
sources. We believe that this is unlikely to be due to an increase in extinction, which by removing flux from the mid-IR and adding it in the
far-IR could mimic a lower dust temperature. As discussed earlier, sources with high extinction and deep silicate absorption are missed by our
24\,$\mu$m selection at $z>$1. However, they do not constitute more than a few per cent of
the population (see also Magdis et al. 2010). Moreover, for the high
redshift sources, where the \textit{Herschel} photometry can more accurately constrain
the slope of the mid-IR-to-far-IR continuum, we find that the sample is mainly fit with low
extinction (shallow silicate absorption depth) models. 

Note that although previous studies have shown that IR galaxies colder than local
equivalents exist in abundance at high redshift --- e.g. results from \textit{ISO} (e.g. Rowan-Robinson
et al. 2005), SCUBA (Kov\'acs et al. 2006; Pope et
al. 2006\nocite{Pope06}; Coppin et al. 2008), \textit{Spitzer} 
(e.g. Symeonidis et al. 2008; 2009; Kartaltepe et
al. 2010a\nocite{Kartaltepe10a}; Patel et al. 2011), \textit{BLAST} (Muzzin et
al. 2010\nocite{Muzzin10}), \textit{Herschel} (e.g. Hwang
et al. 2010, Rowan-Robinson et al. 2010; Smith et al. 2012) --- for
the first time we determine that the mean dust
temperature of the IR-luminous population as a whole decreases as a
function of redshift. Moreover, we find that the decrease in temperature is also a
function of infrared luminosity, i.e. the temperatures of more luminous objects show a
stronger decline from the local to the
early Universe. We note that almost all NIRGs, up to $z\sim$0.5, have temperatures
below 35\,K, whereas for LIRGs the local cold ($T<$35\,K) fraction is 60 per cent
increasing to about 90 per cent at $z\sim$0.5. The ULIRGs show the
largest increase in the cold galaxy fraction, from about 5 per
cent at $z<0.1$ to 30 per cent at $z=1-2$. This is interesting as it
implies that LIRGs undergo more modest evolution than ULIRGs, the
former showing a 2-fold increase in the fraction of cold galaxies,
whereas the latter a 6-fold. Moreover, the cold LIRG fraction in the local Universe is about 12 times higher than
the cold ULIRG fraction, however, we see that this difference
decreases to about 2.5 at $z\sim$0.8. Similarly the cold NIRG fraction
in the local Universe is about 50 per cent higher than the cold LIRG
fraction, whereas at $z\sim$0.4 these fractions are about the same. 
This is evidence that cold galaxies become more
dominant at high redshift. However, it also indicates that there might be a
lower limit in the average dust temperature of the IR-luminous
population, at $T \sim$\,25\,K, towards which systems tend. This is not to say
that $T<$25\,K IR-luminous galaxies do not exist, but these would be
at the tail of the temperature distribution.

As discussed earlier, the $L-T$ relation would be
completely flat, e.g. all IR luminous galaxies would have an average
temperature of $\sim$25\,K, if their sizes or dust masses
increased in proportion to their total IR luminosities (see also Fig. \ref{fig:temperature_herschel}). As it stands,
this is not the case, and so the increased luminosity succeeds in
heating up the dust to a higher temperature. However, the
decrease in average dust temperature with redshift suggests that high redshift
LIRGs/ULIRGs must have more extended IR emitting regions and/or higher
dust masses relative to their lower redshift counterparts, causing the $L-T$ relation at high redshift to become flatter than the
local one. Described phenomenologically, we observe that the temperature evolution of IR-luminous galaxies is
more rapid if their local temperatures are much higher than
25\,K, such as for ULIRGs, than if they are closer to 25\,K such as for
NIRGs. Our results are in agreement with the work presented in Dunne et al. (2011) who find strong evolution in the dust mass density, proposing that IR-luminous
galaxies are dustier at $z\sim$0.5 compared to today, corresponding to a factor 4-5 increase in the dust masses of the most 
massive galaxies. Moreover, CO measurements support the idea of
extended instead of compact star-formation in high redshift ULIRGs,
which have $>>$kpc CO sizes, in contrast to local
equivalent sources which are more concentrated ($<$1kpc) (e.g. Tacconi et
al. 2006\nocite{Tacconi06}, Iono et al. 2009\nocite{Iono09};
Rujopakarn et al. 2011\nocite{Rujopakarn11}). Moreover, significantly higher gas fractions in $z\sim1$ disc
galaxies than in nearby discs have been reported (e.g. Tacconi et al. 2010).
Our results also agree with Seymour et al. (2010), who studied the comoving IR luminosity density (IRLD) as a
function of temperature, proposing that cold galaxies dominate the
IRLD across $0 < z< 1$ and are thus likely to be the main driver behind the increase in
SFR density up to z $\sim$1. 

Although the IR-luminous population and particularly LIRGs and ULIRGs
seem to have more extended dust distribution and/or higher dust masses 
at high redshift, the analysis presented here cannot constrain whether
these systems are characterised by a
merger-induced or isolated star-formation history or where they are
located in the star-formation rate-stellar mass parameter space (e.g. Reddy et
al. 2006\nocite{Reddy06}, 2010\nocite{Reddy10}; Wuyts et al. 2011\nocite{Wuyts11}; Whitaker et al. 2012\nocite{Whitaker12}; Zahid et al. 2013\nocite{Zahid13}). 
Morphological studies of IR-luminous galaxies have presented
contrasting results --- e.g. Bell et al. 2005; Lotz et al. 2006
report that more than half of LIRGs at z$>$ 0.7 are gas-rich isolated
spirals, whereas other studies (e.g. Le Fevre et al. 2000, Cassata et
al. 2005, Bridge et al. 2007; de Ravel et al. 2009) claim an increase in the merger rate out to $z\sim$1
suggesting that more than half of IR luminosity density out to $z\sim$1
is a result of some merger event. Moreover, Zamojski et al. (2011) and
Kartaltepe et al. (2012) find more than 70 per cent of ULIRGs up to $z\sim$2 to be mergers.
These findings are hard to
reconcile and recent evidence from Lotz et al. (2011) shows that
these differences might simply be the result of increasing gas
fractions at high redshift, as the timescale for observing a galaxy to
be asymmetric increases in
tandem with the gas fraction, with the resulting
dust obscuration also being a key factor. With the work presented
here, we are not able to test this argument nor examine the morphological evolution (if
any) of IR galaxies. Nevertheless, our description of the IR-luminous population is
independent of morphological classification. Our results
indicate that the gas-rich environment in the early
Universe might have set or enabled different initial conditions in these systems,
resulting in the observed differences between IR galaxies at high
redshift and their local counterparts. 
The increase in cold ($T<$35\,K) galaxy fraction reported here suggests a greater diversity in
the IR population at high redshift, particularly for (U)LIRGs. In
contrast, the dust properties of the local (U)LIRG population are more uniform and as a result they are not archetypal of
the (U)LIRG population as a whole.

\section {Summary $\&$ Conclusions}
\label{sec:conclusions}

We have examined the dust properties and infrared SEDs of a sample of
1159 infrared-selected galaxies at $z$=0.1-2, using data from \textit{Herschel}/PACS and SPIRE and
\textit{Spitzer}/MIPS (24\,$\mu$m) in the COSMOS and GOODS fields. The unique angle of this work has
been the rigorous analysis of survey selection effects enabling us
work within a framework almost entirely free of selection biases. The results we thus
report should be considered as representative of the aggregate properties of the
star-formation-dominated, IR-luminous ($L_{\rm IR}$$>$10$^{10}$) population up to $z\sim$2. 

We conclude that:
\begin{itemize}
\item{IR-luminous galaxies have mean dust
temperatures between 25 and 45\,K, with $T<25$ and $T>45$\,K sources
being rare. They are characterised by broad-peaked and cool/extended,
rather than warm/compact SEDs, however very cold
cirrus-dominated SEDs are rare occurrences, with most sources having
SED types between those of warm starbursts such as M82 and cool spirals
such as M51. }
\item{The IR luminous population follows a luminosity-temperature ($L-T$) relation, where the more luminous
    sources have up to 10\,K higher dust temperatures. However, the effect of increased dust
    heating is not solely responsible for shape of the $L-T$ relation. We
    find that the increase in dust mass and/or extent of dust distribution of the
    more luminous sources dilutes the increased dust heating,
    flattening the $L-T$ relation and driving it towards the limiting
    scenario of $L \propto R^2$ or $L \propto M_{\rm dust}$ with $T$=const.}
\item{High redshift IR-luminous galaxies are on average
    colder with a temperature difference that increases as a function
    of total infrared luminosity and reaches a maximum of 10\,K. For
    the more luminous (log\,[$L_{\rm IR}$/$L_{\odot}$]$\gtrsim$11.5) sources,
    the $L-T$ relation is flatter at high redshift than in the local
    Universe, suggesting an increase in the sizes and/or dust-masses of
    these systems compared to their local counterparts. }
\item{The fraction of T$<$35\,K galaxies increases with
    redshift as a function of total
    infrared luminosity. Although NIRG fractions are consistent in
    the local and high redshift Universe, LIRGs show a 2-fold increase and
ULIRGs a 6-fold increase. This suggests a greater diversity in the
IR-luminous population at high redshift, particularly for
ULIRGs. }
\end{itemize}


\section*{Acknowledgments}%
This paper uses data from \textit{Herschel}'s photometers SPIRE
and PACS. SPIRE has been developed by a consortium of institutes led
by Cardiff Univ. (UK) and including: Univ. Lethbridge (Canada);
NAOC (China); CEA, LAM (France); IFSI, Univ. Padua (Italy);
IAC (Spain); Stockholm Observatory (Sweden); Imperial College
London, RAL, UCL-MSSL, UKATC, Univ. Sussex (UK); and Caltech,
JPL, NHSC, Univ. Colorado (USA). This development has been
supported by national funding agencies: CSA (Canada); NAOC
(China); CEA, CNES, CNRS (France); ASI (Italy); MCINN (Spain);
SNSB (Sweden); STFC, UKSA (UK); and NASA (USA). PACS has been
developed by a consortium of institutes led by MPE (Germany) and including UVIE (Austria); KU Leuven, CSL, IMEC (Belgium);
CEA, LAM (France); MPIA (Germany); INAF-IFSI/OAA/OAP/OAT, LENS,
SISSA (Italy); IAC (Spain). This development has been supported by the funding
agencies BMVIT (Austria), ESA-PRODEX (Belgium), CEA/CNES (France),
DLR (Germany), ASI/INAF (Italy), and CICYT/MCYT (Spain).

\bibliographystyle{mn2e}
\bibliography{references}

\clearpage
\appendix

\section{Redshifts}
\label{appendixA}
Here we examine the reliability of the photometric redshift catalogues used in this work (see section
\ref{sec:redshifts}) for sources which satisfy the initial criterion for the
selection of our sample (section \ref{sec:initial_selection}): 24\,$\mu$m sources
that have detections (at least 3$\sigma$) at [100 and
160\,$\mu$m] OR [160 and 250\,$\mu$m]. 

\begin{figure}
\centering
\epsfig{file=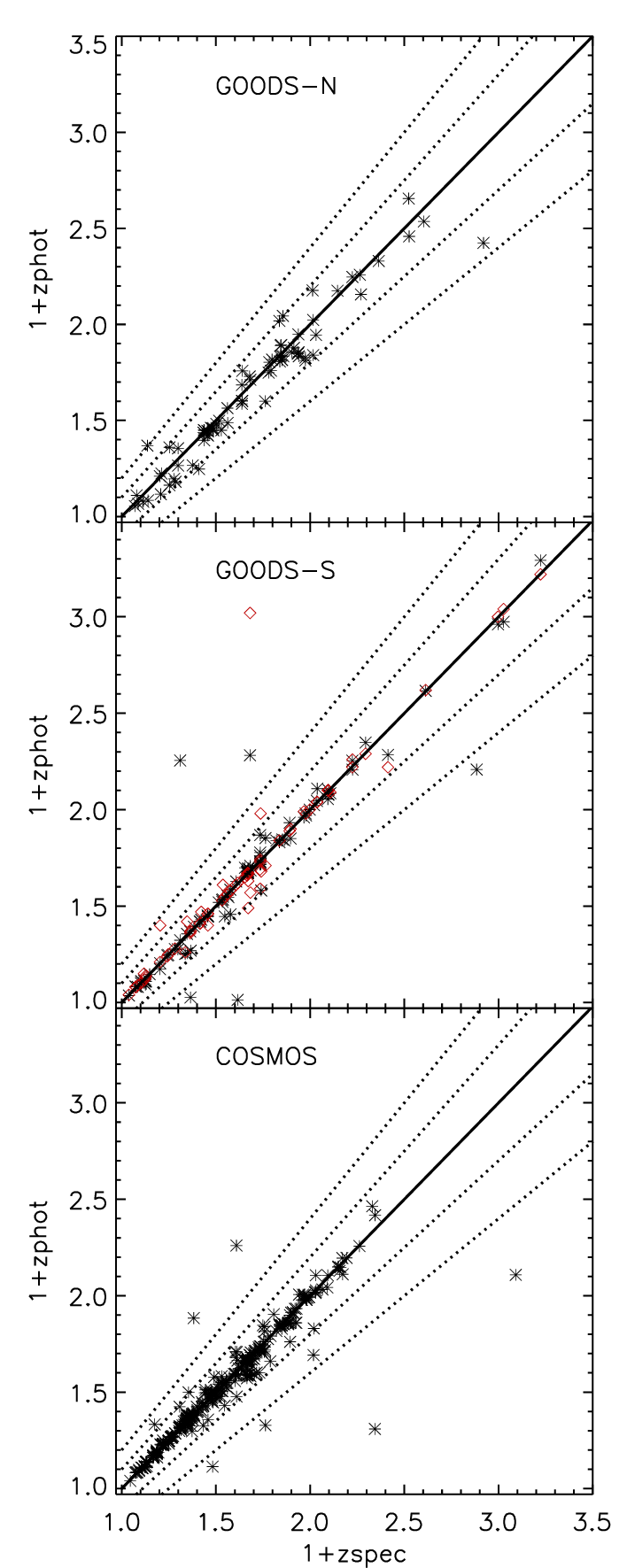, width=0.9\linewidth}
\caption{Comparison of spectroscopic and photometric redshifts for
  sources in the GOODS-N, GOODS-S and COSMOS samples which satisfy the initial criterion for the
selection of our sample (section \ref{sec:initial_selection}): 24\,$\mu$m sources
that have detections (at least 3$\sigma$) at [100 and
160\,$\mu$m] OR [160 and 250\,$\mu$m]. The dashed lines are the 10 per cent boundaries and the dotted lines are the 20 per
cent boundaries. In the middle panel, the asterisks are the Cardamone
et al. (2010) photometric redshifts and the diamonds the
 Santini et al. (2009) photometric redshifts.} 
\label{fig:appendixA_fig1}
\end{figure}

For GOODS-N, we use the redshift catalogue of Berta et al. (2011) which includes
spectroscopic redshifts assembled from various sources
and photometric redshifts compiled using the EAzY code (Brammer et
al. 2008) and up to 14 photometric bands. Berta et al. (2011) note that the fraction of outliers, defined as objects
having $z_{\rm spec}$-$z_{\rm phot}$/(1+$z_{\rm spec}$)$>$0.2, is
about 2 per cent for sources with a
PACS detection. For more details we refer the reader to Berta et
al. (2011). In Fig. \ref{fig:appendixA_fig1} (top-panel) we compare the
photometric and spectroscopic redshifts for the GOODS-N sample, indicating the 10 and 20 per cent uncertainty
regions. There is excellent agreement in the redshifts.

For GOODS-S, we use the MUSYC redshift catalogue of Cardamone et
al. (2010), complemented with the MUSIC redshift catalogue of Santini et
al. (2009). Most redshifts come from Cardamone et
al. (2010) with the addition of some from Santini
et al. (2009). The Cardamone et al. (2010) photometric
redshifts are compiled using up to 28 photometric bands and the EAzY code (Brammer et
al. 2008), resulting to high accuracy $z_{\rm spec}$-$z_{\rm phot}$/(1
+ $z_{\rm spec}$) $\sim$0.00822 out to high redshift. For more details we refer the reader to
the Cardamone et al. (2010) paper. Fig. \ref{fig:appendixA_fig1}
(middle panel) shows a
comparison of photometric to spectroscopic redshifts for the sources
in the GOODS-S sample for both Santini et al. (2009) and Cardamone et
al. (2010) catalogues. There is excellent agreement between the
spectroscopic and photometric redshift in both catalogues, with the
catastrophic failures, which we define as a more than 20 per cent
offset in (1+$z$), being $<$1 per cent of
the spectroscopic sample. 

For COSMOS the spectroscopic redshifts are from Lilly et al. (2009)
and the photometric redshifts from Ilbert et al. (2008), derived using
up to 30 photometric bands. The estimated accuracy of a median
$z_{\rm spec}$-$z_{\rm phot}$/(1 + $z_{\rm spec}$)=0.007 for the galaxies brighter than
i = 22.5. For the sources which host X-ray detected AGN we substitute
the Ilbert et al. redshift with a photometric redshift from Salvato et al. (2009);
(2011). These are derived with a combination of AGN and galaxy
templates and are hence more accurate for galaxies hosting AGN. 
Fig. \ref{fig:appendixA_fig1} (lower panel) shows a comparison of photometric to
spectroscopic redshifts for the COSMOS sample
sources. The agreement is again excellent, with only about 1 per cent
of sources outside the 20 per cent uncertainty region. 

\begin{figure}
\centering
\epsfig{file=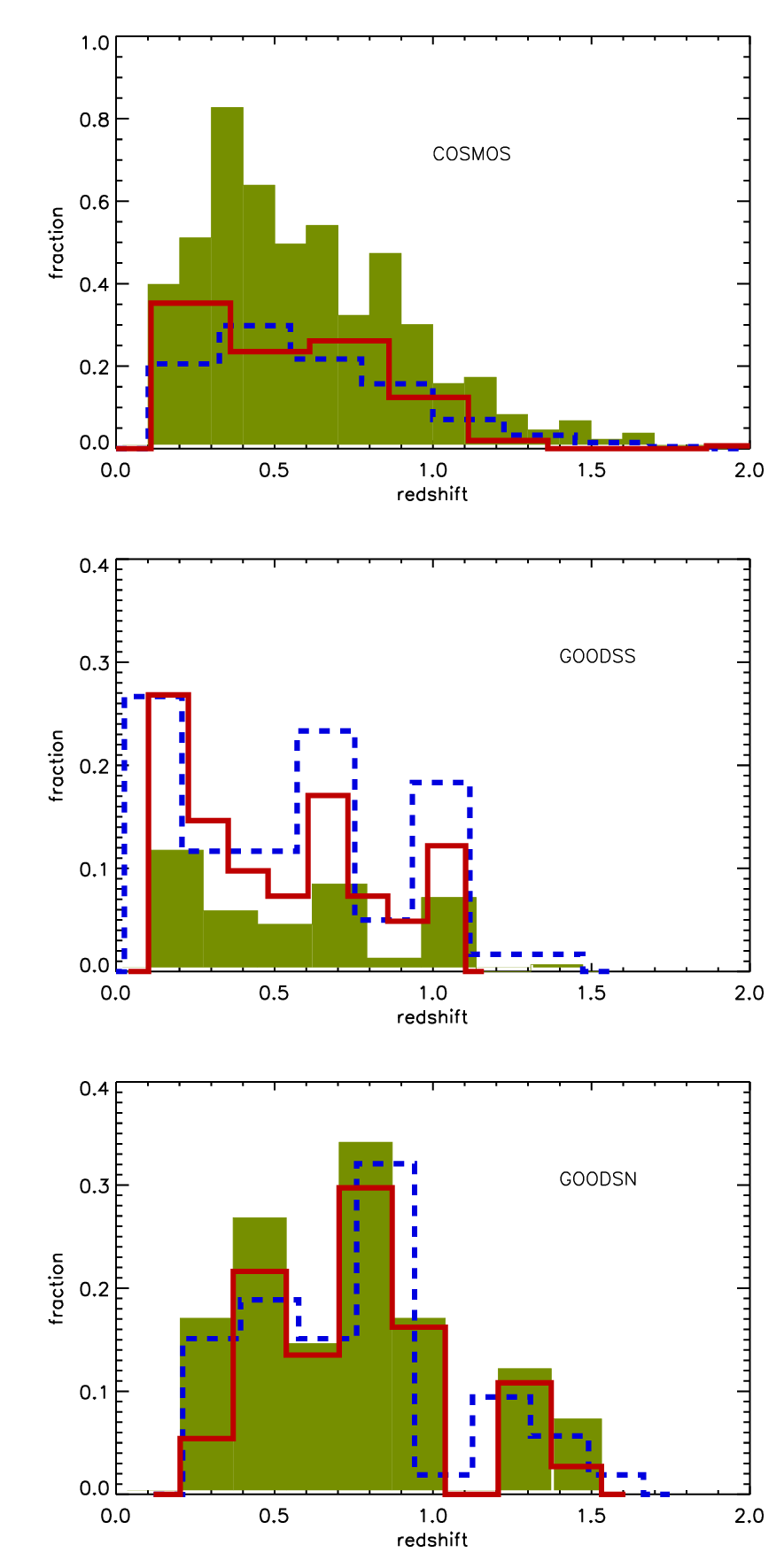, width=0.9\linewidth}
\caption{The redshift distribution of the final \textit{Herschel}
  sample used in this work in COSMOS, GOODS-S and GOODS-N. The filled in
  histogram shows the total redshift distribution whereas the red
  histogram represents the spectroscopic redshifts and the blue
  dashed histogram the photometric redshifts. }
\label{fig:appendixA_fig2}
\end{figure}

\begin{figure}
\centering
\epsfig{file=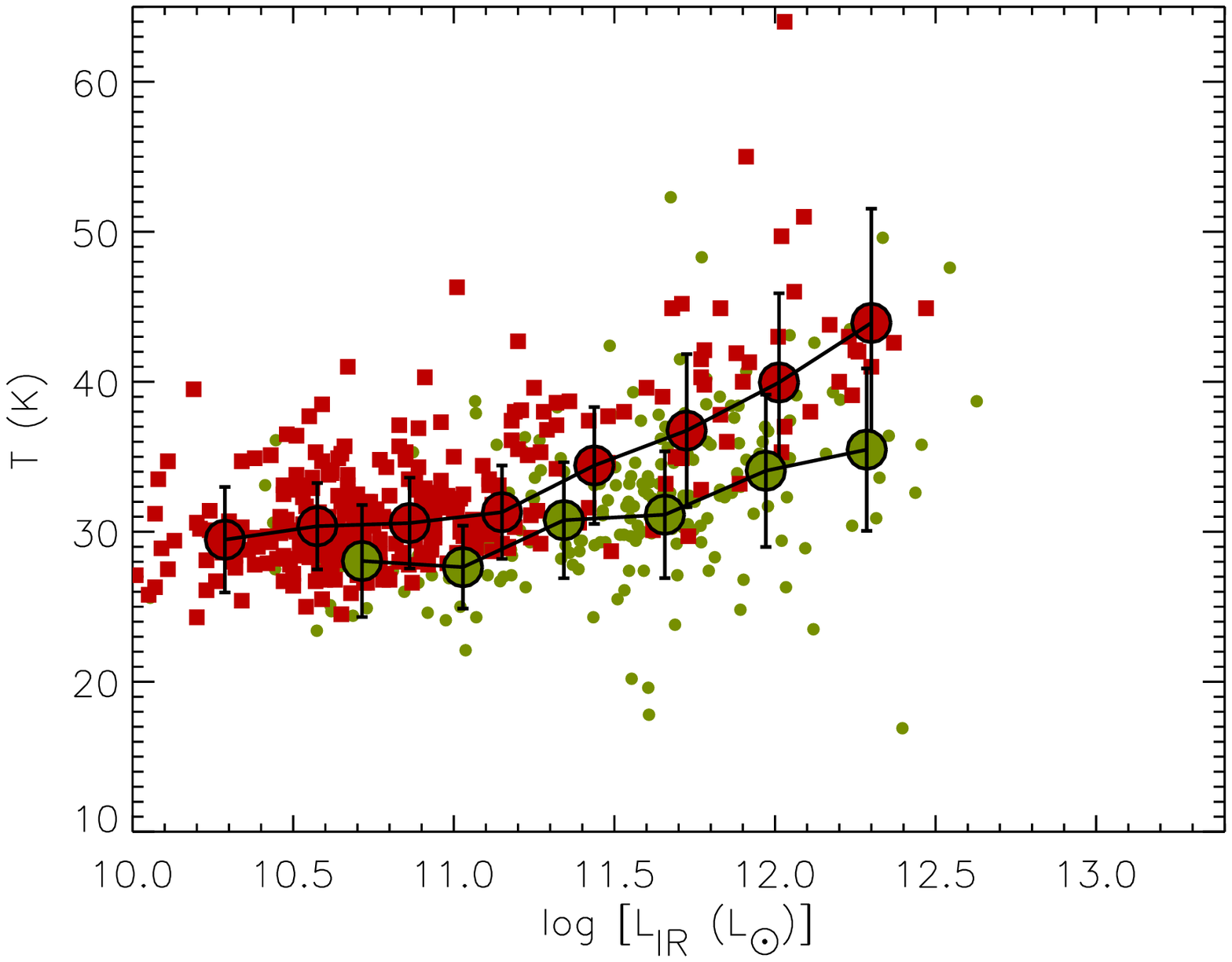, width=0.9\linewidth}
\caption{This figure is reproduced from the main part of the paper, but
  solely with spectroscopic redshifts. We see that the overall results
remain the same, however the statistics are reduced. }
\label{fig:appendixA_fig3}
\end{figure}

We next examine the redshift distribution of our final
\textit{Herschel} sample used in this work assembled in section
\ref{sec:selection}. This is shown in \ref{fig:appendixA_fig2} for the
3 fields. Note that the photometric and spectroscopic redshift
distributions agree, although the photometric redshift distribution
tails off to higher redshifts. We examine whether the use of
photometric redshifts would have any
effect on our results, by reproducing one of our main figures using only
spectroscopic redshifts (Fig. \ref{fig:appendixA_fig3}). We
see that using only spectroscopic redshifts does not change the
overall differences we find between the \textit{Herschel} and local
sample, however it does significantly reduce the statistics.

\label{lastpage}

\end{document}